\documentclass[useAMS,usenatbib]{mn2e}
\usepackage{graphicx}
\usepackage{natbib}
\usepackage{times}
\usepackage{float}
\usepackage{hyperref}
\usepackage{amsmath}
\usepackage{amssymb}
\usepackage{epsfig}
\usepackage{xcolor}
\usepackage{subfigure} 
\definecolor{red}{cmyk}{0,0,0,1}
\definecolor{blue}{cmyk}{0,0,0,1}
\newcommand{\bmth}[1]{\mbox{\boldmath${#1}$}}

\title[Tidal super-harmonic wave excitation in stars]
{Nonlinear tidal excitation of super-harmonic gravity waves in main-sequence stars 
in binary and exoplanetary systems}
\author[ P. B. Ivanov, S. V. Chernov, A. J. Barker ]{P. B. Ivanov$^{1}$\thanks{E-mail:
pbi20@cam.ac.uk (PBI)}, S. V. Chernov $^{1}$\thanks{E-mail:
chernov@td.lpi.ru (SVCh)}, A. J. Barker $^{2}$\thanks{E-mail: A.J.Barker@leeds.ac.uk (AJB)}    \\
$^{1}$Astro Space Centre, P.N. Lebedev Physical Institute, 84/32
Profsoyuznaya Street, Moscow, 117997, Russia  \\
$^{2}$ Department of Applied Mathematics, School of Mathematics, University of Leeds, Leeds, LS2 9JT, UK}

\begin{document}

\date{Accepted. Received; in original form}

\pagerange{\pageref{firstpage}--\pageref{lastpage}} \pubyear{2022}

\maketitle

\label{firstpage}

\begin{abstract}
We study the role of nonlinear effects on tidally-excited internal gravity waves in stellar radiation zones in exoplanetary or binary systems. We are partly motivated to study tides due to massive short-period hot Jupiters, which preferentially orbit stars with convective cores, for which wave breaking near the stellar centre cannot operate. We develop a theory (and test it with numerical calculations) for the nonlinear excitation of super-harmonic ``secondary" waves (with frequencies $2\omega_p$) by a ``primary" tidal wave (with frequency $\omega_p$) near the interface between the radiation zone and convective envelope. These waves have the same horizontal phase speeds to leading order, and this nonlinear effect could contribute importantly to tidal dissipation if the secondary waves can efficiently damp the primary. We derive criteria involving the orbital and stellar parameters required to excite these secondary waves to large amplitudes using a local model of the radiative/convective interface, which we convert to apply to tides in a spherical star. We numerically evaluate the critical amplitudes required for this new nonlinear effect to become important using stellar models, comparing them to the ``conventional" criteria for wave breaking in radiative cores and the application of WKBJ theory near convective cores. The criteria for this new effect are easier to satisfy than the conventional measures of nonlinearity in $1.4$ and $2M_\odot$ stars on the main-sequence. We predict nonlinear effects to be important even for planetary-mass companions around the latter, but this effect is probably less important in stars with radiative cores. 
\end{abstract} 


\section{Introduction}

Tidal interactions play an important role in exoplanetary and close binary systems. Although development of the modern
theory of tides in celestial bodies began over a century ago \citep[starting with e.g.][]{Darwin1880} there are still many open
problems in this area. Perhaps the most important questions concern the efficiency of tidal energy dissipation in stars and giant planets, which are wholly or partly fluid bodies. This is because tidal dissipation can drive evolution of the spins and orbits of stars and planets in close binary and in exoplanetary systems \citep[see e.g. the reviews by][]{O14,Mathis2019}. It has long been argued that when the standard equilibrium (also referred to as quasi-stationary) tides \citep[e.g.][]{Hut1981} are considered in convection zones the dissipative efficiency could be negligibly small in many applications involving planets and main-sequence stars \citep[e.g.][who built upon earlier ideas by \citealt{Zahn66}]{GN77}. Modern numerical calculations have confirmed this qualitative statement\footnote{Unless the mechanism considered by \cite{T21} can work as efficiently as they estimate, though this is highly uncertain \citep{BA21}.}, although the detailed physical picture of the processes involved may be different from what was originally envisaged \citep[see e.g.][]{OL12,DBJ20a,DBJ20b,VB20a,VB20b}. Thus, the equilibrium (quasi-stationary) tides are unlikely to explain the observed parameters of main-sequence close binary and exoplanetary systems, including the orbital evolution inferred in some hot Jupiter systems. However, equilibrium tides are still likely to be the dominant mechanism in giant stars \citep[e.g.][]{VP95,MV12}. Another potentially important tidal dissipation mechanism is dynamical tides \citep[e.g.][]{Cowling1941,Zahn77}, which involve the resonant excitation of low frequency normal modes (or waves) of a planet or a star by the time-dependent tidal potential. Usually, the dynamical tide is thought to consist of internal gravity waves in stably-stratified (radiative) regions, and inertial waves in convective regions of rotating bodies.

In this paper, we consider non-rotating main-sequence stars, or sufficiently slowly rotating ones for which all relevant tidal frequencies are larger than the Coriolis frequency (like in most hot Jupiter hosts), therefore tidal forcing cannot directly excite inertial waves.
On the other hand, internal gravity waves existing in stably-stratified radiative zones may be resonantly excited by tidal forcing, and, in principle, their dissipation can explain certain observational phenomena \citep[including WASP-12 b's inferred orbital decay e.g.][]{M2016,ChPI2017,W17,Patra2020,Yee2020,Barker2020} provided that these waves are sufficiently strongly dissipated inside the star. The regime in which this is the case has been referred to as ``moderately large dissipation" \citep[MLD, by e.g.][]{IPCh,ChPI2017} or as the ``fully damped" or "travelling wave" regime \citep[by e.g.][]{BO2010,Barker2020}. Qualitatively, this regime is expected when the
propagation time of a tidally-excited gravity wave packet is larger than its damping time\footnote{Since the group speed of 
a high radial order gravity wave is much smaller than its phase speed, its propagation time can be much larger than
the dynamical time, so the damping rate (and hence the viscosity or radiative diffusivity damping the wave) need only be `moderately' large.}. However, the `standard' linear mechanisms to damp gravity waves (radiative diffusion and convective damping of the evanescent tails in the envelope) are usually not efficient enough to justify the validity of this regime for close orbits \citep[e.g.][]{T98,GD98}. In this situation various non-linear effects have been proposed for which this fully damped (MLD) regime may be possible, such as weakly non-linear mode-mode interactions \citep[e.g.][]{BO2011,W12,EW16}, or wave breaking near the centres of solar-like stars possessing radiative cores in which the waves can be geometrically focused and attain large amplitudes \citep{GD98,OL2007,BO2010,Barker2011,Barker2020}. Note that, although different theoretical approaches (for example, the normal mode formalism of \citealt{IPCh} or the low-frequency asymptotic approach taken by e.g.~\citealt{GD98}) should, in principle, give the same results for, say, the conditions under which non-linear effects are predicted to be important, in practice they use various different simplifying assumptions, and could therefore differ in their quantitative predictions. Hence, it is worthwhile reproducing results obtained in one formalism with those of another when possible.

Wave breaking near the centres of solar-type stars with radiative cores can lead to efficient wave absorption through the formation a critical layer \citep[e.g.][]{BO2010,Barker2011}, which can naturally explain the occurrence of the fully damped/MLD regime. For example, it has been proposed that WASP-12 has a radiative core due to being a subgiant, for which this mechanism may explain the inferred orbital evolution \citep[e.g.][]{W17}. In the current Sun, the criterion for the onset of wave breaking requires that a planet in a one-day orbit must exceed about 3 Jupiter masses, though this mass threshold is a strong function of the stellar mass and age \citep{BO2010,Barker2020}. At the end of the main-sequence, much lower mass planets can cause wave breaking in the stellar core and potentially be destroyed. However, stars with masses even slightly larger than 1.1 $M_\odot$ typically have convective cores on the main-sequence. Although a similar criterion for wave breaking can be formulated in the radiation zones of these stars using WKBJ theory \citep[e.g.][]{S18,Barker2020}, the threshold companion masses to cause wave breaking are much larger than the ones obtained in solar-like stars, and this effect appears unlikely to ever be important for planetary-mass companions until the star evolves off the main sequence. It is important to verify the validity of this result, and to explore whether tides in F-type stars can be in this fully damped regime or not, because many of the most massive ultra-short period hot Jupiters have been observed to orbit such stars \citep[such as WASP-18b,][]{Wilkins17}.

The criterion for wave breaking is equivalent to the condition that the radial gradient of the total specific entropy, including both the radiative background and the wave, becomes negative. This condition is always formally satisfied in the radiation zone at a point sufficiently close to the interface between the inner radiative region and an outer convective envelope (as we will show in Section \ref{simple}). This suggests that non-linear effects acting on tidal perturbations could be important in this region of a star even though this is not predicted from WKBJ theory. In this paper, we analyse the nonlinear dynamics of these perturbations in some detail by considering a region of small radial extent near such an interface. We use weakly nonlinear theory (second order perturbation theory) to find a condition for which the amplitudes of the generated secondary waves by the primary tidal waves become comparable with the amplitude of the latter. We apply the formalism of \cite{IPCh} assuming the fully damped/MLD regime to describe the primary tidal wave (i.e. the first order perturbations). Additionally, for simplicity, we assume that, after certain modifications, results obtained in planar geometry can be applied to a spherical star. We also adopt the Boussinesq approximation for the equations of motion \citep[e.g.][]{SV1960}, and assume that square of the Brunt-V\"{a}is\"{a}l\"{a} frequency in the radiation zone depends linearly on the distance to the interface provided that this distance is small \citep[see e.g.][for a discussion of this point]{Barker2011,IPCh}.

We find, in agreement with results recently obtained in fluid dynamics and oceanography \citep[see e.g.][]{Wunsch2017,Baker2020}, that first order tidal perturbations generate (through their nonlinear self-interaction) super-harmonic second order perturbations with approximately double the frequency and wavenumber (and hence the same horizontal phase speed). We derive the conditions required for the secondary super-harmonic waves to attain approximately the same amplitudes as their primary waves in our Cartesian model, and then apply this to realistic stellar models assuming that the amplitude of the primary tidal wave (at first order) is in the fully damped/MLD regime. Our criterion is written in terms of the quantity $q/(1+q)$, where $q$ is the mass ratio (secondary perturber mass/primary star mass), which must be larger than a certain critical value, $C_{crit}$, which is a function of the orbital period and stellar parameters. We apply this criterion to a set of main-sequence stellar models with masses $M=1$, $1.4$ and $2M_{\odot}$ with different ages. 

We find that the criterion for nonlinear self-interaction of the primary tidal waves to be important in generating super-harmonic secondary waves near radiative-convective interfaces can be satisfied much more easily than the (WKBJ) criterion for wave breaking near convective cores, in most of our models with $M > M_{\odot}$. It would be interesting to explore with future numerical simulations whether tidally-excited primary waves that satisfy our amplitude criterion could be damped efficiently enough by this mechanism to validate the occurrence of the fully damped/MLD regime in these stars \citep[as assumed by e.g.][]{ChPI2017,Barker2020}. If so, then our results may have important implications both for hot Jupiter and close binary systems. In particular, the occurrence of this fully damped/MLD regime allows straightforward prediction of orbital decay rates in hot Jupiter systems, and orbital evolution in some close binary systems, as long as the properties of the stars and orbital properties are known.    

The structure of this paper is as follows. In \S~\ref{dense}, we recap the main results required to apply the formalism of \citet{IPCh} to obtain the linear tidal response. We then provide simple estimates to predict when nonlinearity might be expected to be important for tidally-excited gravity waves in \S~\ref{simple} at three particular locations: near the centres of radiative cores, at the radiative interface with a convective core, and at the interface between a radiation zone and a convective envelope. This motivates the more detailed calculations of the generation of super-harmonic gravity waves by weakly nonlinear interactions in a local Cartesian Boussinesq model of the transition region between a radiation zone and a convective envelope in \S~\ref{boussi} and \ref{weaklynonlin}. We confirm our analytical results by comparing them with numerical calculations in \S~\ref{numerical}. We then derive simple criteria to predict when this new nonlinear effect is likely to become important in \S~\ref{relation} and then apply these to stellar models in \ref{application}. Finally, we present our conclusions and a discussion in \S~\ref{conclusions}. { In the main text we assume that the star is non-rotating. However, we briefly discuss the most important correction to our results if the star rotates slowly (which is the appropriate regime for many stars hosting short-period planets) in Appendix \ref{app}}.
        
\section{Linear tidal response in radiation zones}
\label{dense}

We adopt the following notation and conventions throughout this paper, unless specified otherwise: $M_*$ and $R_*$ are the stellar mass and radius, respectively, $G$ is the gravitational constant, $M_p$ is the perturber's mass, and the mass ratio is $q=M_p/M_*$. We define
$\Omega_*=\sqrt{GM_*/R_*^3}$ (i.e. the dynamical frequency) and the mean density $\bar \rho=3M_*/(4\pi R_*^3)$. We express all quantities of interest in these natural units. Namely, the density $\rho$ is represented as $\rho=\bar \rho \tilde \rho$, 
the Lagrangian displacement vector ${\boldsymbol \xi}=R_* \tilde {\boldsymbol \xi}$, and all quantities 
having the dimension of frequency are expressed in units of $\Omega_*$. To distinguish dimensional and dimensionless 
quantities we assign tildes to the latter. 

We follow closely \citet[][hereafter IPCh]{IPCh} and assume that all quantities of interest may be represented 
as a discrete Fourier series in time and over the azimuthal angle $\phi$. In particular, the Lagrangian displacement
vector induced at a point in the star by tidal interactions, ${\boldsymbol \xi}$, can be represented as
\begin{equation}
{\bmth \xi}=\sum_{m,k}{\bmth \xi}_{m,k}e^{-i\omega_{m,k}t+im\phi}+c.c.,
\label{l1}
\end{equation}
where $c.c.$ denotes the complex conjugate, it is implied that summation over the azimuthal number $m$
contains only terms with $m=0$ and $2$, and the forcing frequency 
\begin{eqnarray}
\omega_{m,k}=k\Omega_{orb}-m\Omega_{r},
\label{l2}
\end{eqnarray}
where $\Omega_{orb}$ and $\Omega_{r}$ are the orbital frequency and spin angular velocity of the star, respectively, with $k$ being an integer \footnote{{ Note that equation (\ref{l2}) implies that $\omega_{m,k}$ are defined in the rotating frame. In the inertial frame 
we have $\omega_{m,k}=k\Omega_{orb}$.}}.
Hereafter we assume that the star is effectively non-rotating by setting $\Omega_{r}=0$ {in the main text, and briefly discuss the main effects caused by a slow stellar rotation in Appendix \ref{app}.}.

IPCh derived expressions for ${\bmth \xi}_{m,k}$ in terms of the eigenvectors of free stellar
pulsations, ${\bmth \xi}_j$, as 
\begin{equation}
{\bmth \xi}_{m,k}={A_{m,k}\over 2}\sum_{j}{Q_j\over \omega_{m,k}n_j(\omega_{m,k}-\omega_j+i\omega_{\nu})}{\bmth \xi}_j,
\label{l3}
\end{equation}  
where $j$ is an integer, which represents a sum over all of the free modes of the star.
Here $A_{m,k}$ are quantities characterising the amplitudes of the Fourier components of the tidal potential, which are given explicitly in Appendix A of IPCh, for example. For our purposes, we only need to know $A_{2,2}$, which is given by equation (\ref{e2n}) below. The quantities $Q_j$, $n_j$ and $\omega_j$  are the so-called tidal overlap integral (determining how efficiently a given mode is excited by the tidal potential), norm and eigenfrequency of a particular free mode, while $\omega_{\nu}$ is its damping rate. 

For a non-rotating star perturbed by the quadrupolar component of the tidal potential, ${\bmth \xi_j}$ can be expressed in terms of spherical harmonics $Y^{m}_{2}$ as 
\begin{equation}
{\bmth \xi}_{j}=e^{-im\phi}\left\{\xi_j(r)Y^{m}_2(\theta,\phi){\bf e}_r+\xi_{j,S}(r)r\nabla Y^m_2(\theta,\phi))\right\}, 
\label{l4}
\end{equation}  
where the standard spherical polar coordinate system centred on the star $(r,\theta, \phi)$ is used, ${\bf e}_r$ is the unit vector in the
radial direction, and the plane $\theta=\pi/2$ coincides with the orbital plane. The presence of $e^{-im\phi}$ in front of the brackets in (\ref{l4}) stems from our initial definition (\ref{l1}), in that the dependence of ${\bmth \xi}$ on $\phi$ is already taken into account
there and it is implied that the eigenvectors do not depend on this angle.

For a non-rotating star we have 
\begin{eqnarray}
Q_j&=&2\int_0^{R_*} \rho(r) r^{3} (\xi_j + 3\xi_{j,S}) dr,\\
n_j&=& \int_0^{R_*} \rho(r) r^2 ({\xi_j^2} + 6{\xi_{j,S}^2}) dr,
\label{l5}
\end{eqnarray} 
where $\rho (r)$ is the stellar density. Note that, for clarity, we explicitly use the dimensional definition for $Q_j$ and $n_j$ here, contrary to e.g. IPCh and references therein.

The summation in the expression for ${\bmth \xi}_{m,k}$ can be approximately performed 
under the assumption that the spectrum of eigenmodes is dense and regular. This is appropriate for the case of high-order $g$-modes, for example. In this case only modes with frequencies approximately equal to $\omega_{m,k}$ contribute
to the sum over $k$ in (\ref{l3}) and the difference is determined by the values of the factor $\omega_{m,k}-\omega_j+i\omega_{\nu}$ in the denominator for different modes, while other quantities can be taken out of
the sum over $k$. Let's assume that $\omega_{m,k}$ is close to a particular eigenfrequency, with an index $j=j_0$, and then
write down $\omega_{j_0}=\omega_{m,k}+\Delta \omega_{j_0}$, where $|\Delta \omega_{j_0}| \ll |\omega_{j_0}|$ is a frequency offset.
Also, we assume that eigenfrequencies with indices $j \sim j_0$ 
depend approximately linearly on the difference $l=j-j_0$: $\omega_j=\omega_{j_0}+(d\omega/dj)l$, where $d\omega/dj$ stands 
for the difference $\omega_{j_0}-\omega_{j_0-1}$. Under these assumptions, the factor $\omega_{m,k}-\omega_j+i\omega_{\nu}\approx
i\omega_{\nu}-(d\omega/dj)l-\Delta \omega_{j_0}$, and we can write down
\begin{equation}
{\bmth \xi}_{m,k}=-{A_{m,k}Q_{j_0}\over 2\omega_{m,k}(d\omega/dj)_{j_0}n_{j_0}}S{\bmth \xi}_{j_0}, 
\quad S=\sum^{\infty}_{l=-\infty}{1\over \delta +l-i\kappa},
\label{l6}
\end{equation} 
where $\delta = {\Delta \omega_{j_{0}} \over d\omega/dj}$ is a dimensionless frequency offset, $\kappa ={\omega_{\nu} \over d\omega/dj}$ is a dimensionless damping rate, we formally extend 
summation to infinite limits and set $\omega_{j_0}=\omega_{m,k}$ in the terms outside the sum. The sum $S$ can be evaluated 
using complex variable theory (residue calculus) with the result
\begin{equation}
S=-\pi \cot(\pi(i\kappa-\delta)).
\label{l7}
\end{equation} 
Of particular importance is the case of so-called `moderately large dissipation' (MLD), which occurs when the damping rate is larger than
the distance between two neighbouring eigenfrequencies i.e. $\kappa > 1$. For our calculations below related to this regime we use $\kappa=1$ for all values of $\delta$. Note that in this case $S(\kappa=1, \delta)$ 
is very close to its limiting value $S(\kappa\rightarrow \infty, \delta)=i\pi$.

\section{Simple estimates for tidally-excited gravity waves to become nonlinear}
\label{simple}

It is often assumed that internal gravity waves break (or become essentially nonlinear) when the maximum magnitude of the (negative part of the) radial gradient of the Eulerian perturbation of the specific entropy, $\partial_r s^\prime$ becomes larger than the background value, $ds/dr$ \citep[see e.g.][]{OL2007,BO2010,BO2011}, since then there is a portion of the wave in which the entropy profile is overturned (i.e. with a decreasing radial gradient which would be ``convectively unstable").  The Lagrangian perturbation of specific entropy, $\Delta s$, is clearly zero for adiabatic perturbations. We note that $\Delta s=s^{\prime}+{ds\over dr} \xi^r$, where  $\xi^r$ is the radial component of ${\bmth \xi}$. Hence, from the condition $|\partial_r s^\prime|>|{ds\over dr}|$,
we readily find
\begin{equation}
\left\vert {{\partial \over \partial r}}\xi^r +{{d^2 s\over dr^2}\over {ds\over dr} }\xi^r\right\vert > 1,
\label{e1}
\end{equation}
where this expression is meant to be evaluated in stellar radiation zones, and we consider its maximal value over 
the coordinates $r,\theta, \phi$, and a tidal forcing period.

In what follows, we consider the simplest case of an approximately circular orbit, in which only the term
with $m=k=2$ is important in the summation in (\ref{l1}), and the relevant tidal potential component is
$A_{2,2}$, as defined in equation (A2) of IPCh, which has the following explicit form
\begin{equation}
A_{2,2}\approx -{1\over 2}\sqrt{{6\pi\over 5}}{GM_p\over a^{3}},
\label{e2n}
\end{equation}
where $a$ is the orbital semi-major axis. We also have $\omega_{2,2}=2\Omega_{orb}$, and from (\ref{l1}) and  (\ref{l6})
we obtain
\begin{equation}
{\bmth \xi} = -{\pi \over 2} A_{2,2} {Q \over {d\omega\over dj}\Omega_{orb} n}{\bmth \xi}_{j_0}\sin(\Psi),
\label{e2}
\end{equation}
where $\Psi=2(\Omega_{orb}t-\phi)$. It is implied below that all quantities associated with an eigenmode correspond to one having an approximately resonant frequency $2\Omega_{orb}$. We do not show the corresponding mode index $j_0$ in expressions for the overlap integral $Q$, the norm $n$, the eigenfrequencies $\omega$, and their differences $d\omega/dj$ hereafter. Since we require that $\xi^r$  in (\ref{e1}) should take its  largest value over a wave period, we set $\sin \Psi =1$ in (\ref{e2}). 

From (\ref{l4}) it follows that the radial component of ${\bmth \xi}_{j_0}$, $\xi^r_{j_0}$, entering (\ref{e1}) is expressed in terms of $\xi(r)$ through the factor $e^{-2i\phi}Y^{2}_{2}(\theta,\phi)$, which is a function of the angle $\theta$ only. This factor takes its largest value at $\theta=\pi/2$, where it is equal to ${1\over 4}\sqrt{{15\over 2\pi}}$. Note that $\xi(r)$ is assumed to be known from the solution of the standard problem of free stellar pulsations. We express $\xi^r_{j_0}$ in terms of $\xi(r)$, then substitute (\ref{e2n}) in (\ref{e2}), to obtain the radial component
\begin{equation}
\xi^r={3\pi\over 16}{q\over 1+q}{\Omega_{orb}\over d\omega/dj}{Q\over n}\xi.
\label{e2nn}
\end{equation}
We then substitute (\ref{e2nn}) into (\ref{e1}) to obtain our criterion for nonlinearity
\begin{equation}
{q\over {1+q}} > C_{crit},
\label{e3}
\end{equation} 
where 
\begin{equation}
C_{crit}={16\over 3\pi}{\left\vert {d \omega\over dj}\right\vert \over \Omega_{orb}}{n\over \left\vert Q \left({d \xi \over dr} +{{d^2 s\over dr^2}\over {ds\over dr}}\xi\right)\right\vert}.
\label{e4}
\end{equation}
Note that (\ref{e4}) should be evaluated in radiative zones, and then its minimal value should be used in (\ref{e3}) to determine whether or not wave breaking (or other important nonlinear effects) is predicted.

\subsection{Wave breaking in the radiative cores of solar-like stars}
\label{WaveBreakCore}

Firstly, we wish to apply the criterion (\ref{e3}) to the case of a solar-type star with a radiative core, under the assumption 
that the minimum value of $C_{crit}$ as a function of $r$ is reached at the centre. This is the case that has been studied by e.g.~\citet{GD98,OL2007,BO2010,Barker2011,W12,Barker2020}.
Following \citet{OL2007} we assume that the Brunt-V\"{a}is\"{a}l\"{a} frequency, $N(r)$, depends linearly on the distance to the centre of the star, i.e. $N=A_{centre}r$. We also assume that the entropy gradient is proportional to $N$ (and $g\sim r$ there also). It then follows from \citet{OL2007} or \citet{Barker2011}, that near the centre $\xi$ can be expressed in terms of Bessel functions as
\begin{equation}
\xi(r)=C_{centre}r^{-3/2}J_{5/2}(R_{centre}r),
\label{e5}
\end{equation}
where $C_{centre}$ and $R_{centre}$ can be found by matching (\ref{e5}) to a WKBJ solution $\xi_{WKBJ}$ valid in the 
radiative region of a solar-like star for a g-mode with a large radial order. Such a solution is given e.g. by equation (88) of IPCh. 
Close to the centre it has the form
\begin{equation}
\xi_{WKBJ}=-C_{WKBJ}{1\over \sqrt{\rho_{centre} A_{centre}}r^2}\sin(\sqrt{6}{A_{centre}r\over \omega}),
\label{e6}
\end{equation}
where $\rho_{centre}$ is the central stellar density and $\omega$ is the mode eigenfrequency. The expression (\ref{e5}) should match
(\ref{e6}) at large values of the argument of the Bessel function, which allows us to obtain
\begin{equation}
R_{centre}=\sqrt{6}{A_{centre}\over \omega}, \;\;\& \;\; C_{centre}=6^{1/4}\sqrt{{\pi\over 2}}{C_{WKBJ}\over \sqrt{\rho_{centre}\omega}}.
\label{e7}
\end{equation} 

We now consider the opposite limit of small values of the argument of (\ref{e5}) to obtain our nonlinearity criterion. Taking into account (\ref{e7}) we obtain
\begin{equation}
\xi \approx {6^{3/2}\over 15}{C_{WKBJ}A_{centre}^{5/2}\over \rho_{centre}^{1/2}\omega^3}r.
\label{e8}
\end{equation} 
Now, substituting (\ref{e8}) into (\ref{e4}) and noting that ${d^2s\over dr^2}$ vanishes under our assumptions, we obtain
\begin{equation}
C_{crit, centre}={80\over 6^{3/2}\pi}{n\over Q}\left\vert\frac{d\omega}{dj}\right\vert{\rho_{centre}^{1/2}\omega^{3}\over C_{WKBJ}A_{centre}^{5/2}}\Omega_{orb}^{-1}
\label{e9}
\end{equation} 

As discussed in IPCh, for the modes described by WKBJ theory, the constant $C_{WKBJ}$ can be expressed through the norm $n$ and eigenfrequency $\omega$ as (see their equation 109)
\begin{equation}
C_{WKBJ}=\sqrt{{2n\over I}}\omega, \;\;\text{where} \quad I=\int_{0}^{r_c}{dr\over r}N,
\label{e10}
\end{equation}  
and where $r_c$ is the radius at the base of the convective envelope, while the frequency difference $d\omega/dj$ is given by their equation (125) with $k=2$ and $\Lambda=6$, i.e.
\begin{equation}
\frac{d\omega}{dj}={4\pi \over \sqrt{6}}{\Omega^2_{orb}\over I}.
\label{e11}
\end{equation}  
We substitute (\ref{e10}) in (\ref{e8}), then the result of this and (\ref{e11}) in (\ref{e9}), and set $\omega=2\Omega_{orb}$, to obtain
\begin{equation}
C_{crit, centre}={320\over 9\sqrt 2}{\rho_{centre}^{1/2}\over I^{1/2}A_{centre}^{5/2}}{1\over \hat Q}\Omega_{orb}^3,
\label{e12}
\end{equation} 
where
\begin{equation}
\hat Q=Q/\sqrt{n},
\label{e12a}
\end{equation} 
is a `normalised' overlap integral whose value does not depend on the eigenfunction amplitudes.

In the low frequency asymptotic limit $\Omega_{orb}\rightarrow 0$, one can show (see IPCh) that under the usual assumption that $N^2$ close to the base of the convective envelope scales linearly with the distance from it, $\hat Q \propto \Omega_{orb}^{17/6}$. Therefore, $C_N\propto \Omega_{orb}^{1/6}$. Comparing with our criterion (\ref{e3}), we can see that this scaling agrees with equations (A7) and (A9) of \citet{OL2007} and with equations (49) and (50) of \citet{Barker2020}. We have also confirmed that our criterion gives quantitatively similar results to those in the literature in the low frequency limit.

\subsection{Wave breaking near convective cores}

Stars with masses exceeding approximately $1.1M_\odot$ (in particular we consider F-type or A-type stars) typically possess both convective envelopes and convective cores with a radiative zone in between. The transition of $N^2$ from its value in the intermediate radiative zone to the convective core is often very abrupt relative to the wavelength of the tidal waves. In this case it is reasonable to assume that the WKBJ approximation remains valid all the way down to the transition radius, $r_{core}$. In such a situation, the criterion for wave breaking is that the maximum value of $|d\xi^r/dr| >1$, where $\xi^r$ is determined by 
(\ref{e2nn}). The maximal value over one oscillation period is considered and $\xi$ entering (\ref{e2nn}) is given by eq. (88) of 
IPCh at $r\rightarrow r_{core}$. We find
\begin{equation}
\left\vert\frac{d\xi^r}{dr}\right\vert_{max}={9\sqrt{2}\over 32}{q\over 1+q}\rho^{-1/2}r^{-5/2}{Q\over n^{1/2}}{\sqrt{N_{core}I}\over \Omega_{orb}},
\label{core1}
\end{equation}
where\footnote{Note that, typically, the dependence of $N^2$ on $r$ has very sharp features in the vicinity
of $r_{core}$ in stellar models. We believe that these features should be discarded when evaluating $N_{core}$, since they are probably unphysical and would be smoothed out by various hydrodynamical mixing processes, including convective overshoot.} $N_{core}=N(r\rightarrow r_{core})$.
Accordingly, from the condition $|d\xi^r/dr|_{max} >1$ we obtain
\begin{equation}
{q\over 1+q} > C_{crit, core}, \quad 
C_{crit, core}={32\over 9\sqrt{2}}\rho^{1/2}r^{5/2}{\Omega_{orb}\over \sqrt{N_{core}I}}{1\over \hat Q},
\label{core2}
\end{equation}
where we note that all quantities in (\ref{core2}) should be evaluated at $r\rightarrow r_{core}$. The approach here is essentially the same as the WKBJ estimate in equations (51) and (53) of \citet{Barker2020} except that the amplitude of the wave is determined by directly computing overlap integrals rather than applying the energy flux obtained by an asymptotic low-frequency analysis (though except for the shortest orbital periods these approaches should give similar results).
 
\subsection{Wave breaking near the transition from a radiation zone to a convective envelope?}

Formally, we can see that $C_{crit}$ can be arbitrary small just below the base of a convective zone. In this region, the 
WKBJ approximation is strictly invalid, and the corresponding wave solution predicts $\xi$ to be approximately constant when ${r_c-r\over r_c} \ll 1$, where $r_c$ is radius at the base of the convective zone. 
On the other hand, the term ${d^2 s \over dr^2}/ {ds\over dr}$ is expected to diverge when ${r_c-r\over r_c} \rightarrow 0$, which leads to $C_{crit}$ formally tending to zero in the same limit. This observation suggests that the situation close to the base of a convective 
envelope deserves a special treatment. This is our aim in the next few sections.  

\section{Local weakly nonlinear analysis of super-harmonic wave generation near the base of a convective envelope}
\label{boussi}

In this section, we consider the evolution of stellar perturbations near the base of a convective envelope and assume that 
the square of the Brunt-V\"{a}is\"{a}l\"{a} frequency has a linear dependence on the distance from the interface at $r_c$, such that
\begin{equation}
N^{2}\approx A_{c}(r_c-r).
\label{e14}
\end{equation}
We adopt units of length here in terms of $r_c$, and units of time in terms of $\omega_c^{-1}$, where $\omega_c=\sqrt{A_cr_c}$. 
We assume that the radial wavelength of gravity waves is small enough that we can approximately describe them using 
Cartesian geometry with local coordinates $(x,y,z)$, such that $z=r_c-r$, and $x,y$ are local horizontal coordinates. Additionally, we employ the well-known Boussinesq approximation for the equations of motion \citep{SV1960}, which is valid for low frequency gravity waves in a region of small spatial extent. In this approximation, the corresponding adiabatic non-linear equations of motion in the radiation zone can be written
\begin{equation}
\dot {\bf U}+({\bf U}\cdot \nabla){\bf U}=-\nabla P + b{\bf e}_z, \quad \nabla \cdot {\bf U}=0, \quad \dot b+({\bf U}\cdot \nabla)b=-zU^{z}.
\label{b1}
\end{equation}
Here a dot stands for an Eulerian time derivative, ${\bf U}$, $P$ and $b$ are perturbations of the velocity and pressure, and the buoyancy variable\footnote{Defined by $b=-g\rho'/\rho_c$, where $\rho'$ is an Eulerian density perturbation, $\rho_c$ is a constant reference density, and $g$ is the local acceleration due to gravity.}, respectively. We have assumed that $N^2$ is given by equation
(\ref{e14}) and that all dynamical variables are dimensionless by being expressed in the units indicated above. 
Note that from (\ref{b1}) it follows that 
\begin{equation}
U^{i}_{,k}U^{k}_{,i}=-\Delta P +{\partial b\over \partial z},
\label{b2}
\end{equation}
where summation over repeated indices is assumed from now on, and $\Delta$ is the Laplacian operator.

We now consider the $z$ component of the equations of motion, differentiate it with respect to time, and apply the Laplacian operator to the resulting expression. Then, using (\ref{b2}) to eliminate $\dot b$, we obtain
\begin{equation}
\Delta \ddot U^z +\Delta_{\perp}(zU^z)=-\Delta_{\perp} ({\bf U}\cdot \nabla)b +\dot T,
\label{b3}
\end{equation}
where $\Delta_{\perp}={\partial^2\over \partial x^2}+{\partial^2\over \partial y^2}$, and  
\begin{equation}
T={\partial \over \partial z} (U^{i}_{,k}U^{k}_{,i})-\Delta ({\bf U}\cdot \nabla)U^z.
\label{b4}
\end{equation}
In a linear analysis, the right hand side of (\ref{b3}) is ignored, which is formally valid if the solution is of infinitesimally small amplitude.
If we substitute the linear solution to compute the terms on the right hand side of (\ref{b3}) we can formulate a second order problem, which is referred to as a weakly nonlinear analysis. Technically, our solution is then reduced to finding solutions of the same linear equation, but with a forcing term determined by the solution to the first order problem. We can then compare the first and second order solutions for $U^z$. The condition that they are of the same order can be considered as a condition for the breakdown of our perturbation theory.

When looking for the linear solution, without loss of generality we can assume that $U^z$ depends only on $z$, $x$ and $t$, and consider $U^{z}=v(z)e^{i(\omega t +kx)}+C.C.$, where $C.C.$ stands for the complex conjugate. Substituting this ansatz into (\ref{b3}) and neglecting the non-linear terms, we see that it reduces to
\begin{equation}
v_{,zz}=(k^2-{k^2\over \omega^2}z)v.
\label{b5a}
\end{equation}
Solutions of (\ref{b5a}) can be represented in terms of Airy functions $Ai(x)$ and $Bi(x)$:
\begin{equation}
v(z)=C_1Ai(-y)+C_2Bi(-y), \quad y={\left({k\over\omega}\right)}^{2/3}(z-\omega^2). 
\label{b5}
\end{equation}

In order to express the nonlinear source term in terms of the linear solution, it follows from the continuity equation that $U^{x}_{,x}=-U^{z}_{,z}$ and, accordingly, in the linear approximation we have $U^{x}={i\over k}(v_{,z}e^{i\phi}-C.C.)$, where $\phi=\omega t+kx$.
Using these relations we obtain
\begin{equation}
U^{i}_{,j}U^{j}_{,i}=2((v^2_{,z}-vv_{,zz})e^{2i\phi}+v_{,z}v^{*}_{,z}+vv^{*}_{,zz}+C.C). 
\label{b7}
\end{equation}
Since only the term proportional to $e^{2i\phi}$, $(v^2_{,z}-vv_{,zz})$, as well as its complex conjugate, depends on time, and accordingly, contributes to the source of second order perturbations, we only consider this term below. Using equation (\ref{b5a}), we obtain ${d\over dz}(v^2_{,z}-vv_{,zz})={k^2\over \omega^2}v^2$. A similar calculations shows that the term $({\bf U}\cdot \nabla)U^z$ entering (\ref{b4}) does not depend on time at second order, and therefore we obtain 
\begin{equation}
\dot T = 4i{k^2\over \omega}v^2e^{2i\phi}, 
\label{b8}
\end{equation}
where we consider only the term proportional to $e^{2i\phi}$. 

Now let us calculate the term  $\Delta_{\perp} ({\bf U}\cdot \nabla)b$ entering (\ref{b3}). From the last equation in (\ref{b1}) it follows that to first order we have
\begin{equation} 
b=i{z\over \omega}(ve^{i\phi}-v^{*}e^{-i\phi}).  
\label{b9}
\end{equation}
Now we take into account that ${\bf U}\cdot \nabla = {i\over k}(v_{,z}e^{i\phi}-v^{*}_{,z}e^{-i\phi}){\partial \over \partial x}+
(ve^{i\phi}+v^{*}e^{-i\phi}){\partial \over \partial z}$ to obtain
\begin{eqnarray} 
\nonumber
({\bf U}\cdot \nabla)b&=&{i\over \omega}(v^2e^{2i\phi}-{v^{*}}^2e^{-2i\phi}),  \\  
\Delta_{\perp} ({\bf U}\cdot \nabla)b&=&-4i{k^2\over \omega}v^2e^{2i\phi}+C.C..
\label{b10}
\end{eqnarray}
Substituting (\ref{b8}) and (\ref{b10}) in (\ref{b3}) and representing
the second order solution in the form 
\begin{equation}
U^z_{(2)}=we^{2i\phi}+C.C.,
\label{form}
\end{equation} 
we arrive at the equation describing second order perturbations:
\begin{equation} 
{d^2\over dz^2}w-4k^2w+{k^2\over \omega^2}zw=-{2ik^2\over \omega^3}v^2.
\label{b11}
\end{equation} 
This equation describes the nonlinear generation of super-harmonics by the self-interaction of the primary wave, which we assume to have been (linearly) tidally forced. In the next section we will determine approximate solutions to this equation.

\section{A weakly non-linear model eigenproblem}
\label{weaklynonlin}

We now consider a model problem focussing on the interface between convective and radiative regions, which may be shown to be relevant for the global perturbations in a star, as we show later in section \ref{relation}. We assume that $U^z=0$ at finite distances $z=z_r$ into the stable/radiative zone ($z>0$), and $z=-z_c$ into the neutral/convective zone ($z <0$), respectively. We calculate the spectrum of eigenmodes corresponding to solutions of eq. (\ref{b5a}) for this problem and consider a particular `primary' mode with frequency $\omega=\omega_0$. The corresponding eigenfunction is assumed to determine the source term on the right hand side of (\ref{b11}). The response of super-harmonic second order waves is again solved through decomposition of the solution over eigenmodes.

\subsection{Linear solution in the neutral zone}
In the neutral zone for $z < 0$, the solution can be expressed in terms of growing and decaying exponentials,
\begin{equation} 
v=C^1_{n}e^{-kz}+C_n^2e^{kz}=C^2_n(e^{kz}-e^{-k(z+2z_c)}), 
\label{mp1}
\end{equation}
where the last equality follows from our condition $U^z(z=-z_c)=0$. From equation (\ref{mp1}) it follows that
\begin{equation} 
v_{,z}(z=0)=\kappa v(z=0), \quad \kappa=k{(1+e^{-2kz_c})\over (1-e^{-2kz_c})}. 
\label{mp2}
\end{equation}
This condition should be matched to the solution in the radiative zone. Note that when $kz_c \ll 1$, $\kappa \approx 1/z_c$ does not
depend on $k$.

\subsection{Linear solution in the radiative zone and the eigenspectrum}

When $z > 0$ we represent the solution (\ref{b5}) in the form
\begin{equation}
v=C\bar v, \quad \bar v=(\cos (\phi)Ai(-y)+\sin (\phi )Bi(-y)),
\label{b20}
\end{equation}
where $y=(z-\omega^2)/\lambda_*$, and $\lambda_*={({\omega\over k})}^{2/3}$. In the limit $z\rightarrow 0$ it follows from (\ref{b20}) that
\begin{eqnarray} 
\nonumber
v&\approx& C\left( {(\cos \phi +\sqrt{3}\sin \phi )\over 3^{2/3}\Gamma (2/3)}-
{(\cos \phi -\sqrt{3}\sin \phi) \over 3^{1/3}\Gamma (1/3)}{\omega^2\over \lambda_*} \right.\\ 
&&\left.\hspace{2cm}+
{(\cos \phi -\sqrt{3}\sin \phi )\over 3^{1/3}\Gamma (1/3)}{z\over \lambda_*}\right).
\label{b23}
\end{eqnarray}
This implies that $v_{,z}\propto {(\cos \phi -\sqrt{3}\sin \phi )\over \lambda_*}$.  This ratio
should be finite in the considered limit $\lambda_*\rightarrow 0$, and, therefore, $\phi$ should be close to $\pi/6$. We
assume that $\phi=\pi/6+\Delta$, where $\Delta$ is small, and substitute (\ref{b23}) in (\ref{mp2}) to obtain
\begin{equation}
\Delta=-{\Gamma (1/3)\over \Gamma (2/3)}{3^{1/6}\over 2}\kappa \lambda_*\left(1-{\Gamma (1/3)\over \Gamma (2/3)}{\kappa\lambda_*\over 2\cdot
3^{1/3}}-\kappa \omega^2\right).  
\label{mp3}
\end{equation}
Note that the last two terms in the brackets are unimportant in the limit $\omega \rightarrow 0$ and are neglected hereafter. 

We assume that $z_r \gg \lambda_*$. In the 
limit $z \gg \lambda_*$, 
\begin{equation}
v\propto \sin \left({2\over 3}\left({(z-\omega^2)\over \lambda_*}\right)^{3/2}+{5\pi\over 12}+\Delta\right),
\end{equation}
and, from the condition $v(z=z_r)=0$, we obtain
\begin{equation}
{2\over 3}{k\over \omega_n}{\left(z_r-\omega_n^2\right)}^{3/2}+\Delta(\omega_n)=\pi \left(n-\frac{5}{12}\right),  
\label{mp4}
\end{equation}
which is an equation for our eigenfrequencies $\omega_n$. 

When $\omega_n$ is small it can be represented as $\omega_n=\omega^0_n+\omega^1_n$, where
\begin{eqnarray}
\nonumber
\omega^{0}_{n}&=&{2\over 3}{kz_r^{3/2}\over \pi (n-5/12)}, \\
 \omega^1_n&=&-{3\omega^0_n\over 2z_r}\left({\omega^0_n}^2+
{\Gamma (1/3)\over \Gamma (2/3)}{3^{1/6}\kappa \over 2 \sqrt{z_r}}{\left({\omega^0_n\over k}\right)}^{5/3}\right).
\label{mp5}
\end{eqnarray} 
Note that from (\ref{mp5}) it follows that when $\omega^0_n$ and $k$ are given by the first expression, then $2\omega^0_n$ 
is determined by the same expression with a doubled wavenumber $2k$. That means that there are eigenfrequencies corresponding to $k$ and $2k$, which differ only by small corrections, and accordingly there is a secondary mode which is in near-resonance with the nonlinear source term due to the primary. 

\subsection{Weakly nonlinear generation of secondary waves}
\label{wnsec}
In a weakly nonlinear regime we assume that there is no back reaction of the second order perturbations on the primary wave. 
In this case, nonlinear terms entering eq. (\ref{b3}) are assumed to be given by the right hand side of 
(\ref{b11}). We represent (\ref{b3}) in the form
\begin{equation}
\Delta \ddot U^z +\Delta_{\perp}(zU^z)=S, \quad S={8ik_p^2\over \omega_p}C^2_p{\bar v_p}^2e^{2i(\omega_pt+k_px)},
\label{mp6a}
\end{equation} 
where we assign the index $p$ to all quantities corresponding to the primary mode, $\omega_p$ and $k_p$ are related by eq. (\ref{mp5}), 
$v$ is given by (\ref{b20}), we note that $\Delta_{\perp}={\partial^2\over \partial x^2}$,
 and we omit the index $n$ from here on in this section. We seek series solutions to (\ref{mp6a}) of the form
\begin{equation}
U^z=\sum_i a_i(t)\phi_i(x,z),
\label{mp6b}
\end{equation} 
where $\phi_i$ are solutions of the eigenproblem
\begin{equation}
\lambda_i \phi_i=\hat A \phi_i, \quad \hat A = \Delta^{-1}\Delta_{\perp}(z\phi_i)=-k_s^2\Delta^{-1}(z\phi_i),
\label{mp6}
\end{equation} 
where $\Delta^{-1}$ is the inverse of the Laplacian operator, $k_s=2k_p$ and $\lambda_i=\omega_i^2$. It is easy to show that the set of eigenfunctions $\phi_i$ are orthogonal with respect to the inner product
\begin{equation}
N_{i,j}=\int z \phi_i\phi_j^{*} dx dz ,
\label{mp7}
\end{equation} 
where integration over $z$ is performed from $0$ to $z_r$ and we assume, for simplicity, periodic boundary conditions in the $x$ direction with a period $x_b \gg k^{-1}$.  In order to prove this, we multiply (\ref{mp7}) by $\lambda^{*}_j$ to obtain
\begin{eqnarray}
\nonumber
\lambda^{*}_jN_{i,j}&=&\int z \phi_i\Delta^{-1}\Delta_{\perp}(z \phi_j^{*}) dx dz \\
\nonumber
&=&\int \Delta_{\perp} (z \phi_i)\Delta^{-1}(z \phi_j^{*})dx dz \\
\nonumber
&=&\int \Delta \Delta^{-1}\Delta_{\perp} (z \phi_i)\Delta^{-1}(z \phi_j^{*})dx dz  \\
\nonumber
&=& \lambda_i\int z \phi_i \phi_j^{*}dx dz \\
\nonumber
&&\hspace{-0.2cm}
+
\lambda_i\int  \nabla (\phi_i\nabla (\Delta^{-1}(z\phi_j^{*})) -\Delta^{-1}(z\phi_j^{*})\nabla \phi_i)dxdz \\
&=&
\lambda_i N_{i,j}
-{\lambda_i\lambda_j^{*}\over k_s^{2}}\int {\partial\over \partial z} (\phi_i\nabla \phi_j^{*}-\phi_j^{*}\nabla \phi_i) dx,
\label{mp8}
\end{eqnarray}
where we use (\ref{mp6}), integration by parts, the facts that our boundary conditions are periodic in the $x$ direction and that $\Delta_{\perp}\phi_i=-k_s^2\phi_i$,
and known properties of the Laplacian. The last term in (\ref{mp8}) is zero due to our condition (\ref{mp2}), and we arrive at
$(\lambda^{*}_j-\lambda_i)N_{i,j}=0$, which proves the statement. In a similar way we can prove that the eigenvalues $\lambda_{i}$ are real and positive. Since $\lambda_i=\omega_{i}^2$, where $\omega_{i}$ are given by eq. (\ref{mp5}) with $n=i$, this has already been shown and we omit the proof.

We substitute (\ref{mp6b}) in (\ref{mp6a}), use (\ref{mp6}), multiply the result by $\phi_{j}^{*}$, and integrate over $x$ and $z$ to obtain
\begin{equation}
\ddot a_j+\lambda_ja_j=-{\lambda_j\over k^2}{S_j\over N_j},
\label{mp9}
\end{equation}
where
\begin{equation}
S_j=\int \phi_j^{*}S dx dz, \quad N_j=N_{j,j}=\int z\phi_j^{*}\phi_j dxdz. 
\label{mp10}
\end{equation}
Remembering that $S \propto e^{2i\omega_pt}$, solutions to (\ref{mp9}) are 
\begin{equation}
a_j={\omega_j^2\over k_s^2(4\omega_p^2-\omega_j^2)}{S_j\over N_j}e^{2i\omega_pt}, 
\label{mp11}
\end{equation}
where we remember that $\lambda_j=\omega^2_j$. 

\subsection{Excitation of the near resonant secondary mode}

As we have mentioned above, there is a secondary mode with $k_s=2k_p$, whose frequency is close to double the frequency of the primary mode, $2\omega_p$, and, accordingly, to the frequency of the source $S$. Both the secondary mode and the source 
have the same zeroth order frequency $2\omega_n^0$, so their difference is due to the corrections
given by $\omega_n^1$ in equation (\ref{mp5}). Equation (\ref{mp11}) tells us that the amplitude of this secondary mode is expected to be 
much larger than other secondary modes due to the factor $4\omega_p^2-\omega_j^2$ in the denominator. Let us calculate $a_j$ corresponding to this mode. For that, we take into account the difference between these frequencies only in the denominator, in all other expressions we neglect the corrections $\omega_n^1$, and assume that the eigenfunction of the secondary mode is given by (\ref{b20}) with $\phi=\pi/6$ multiplied by $e^{2ik_px}$. We also set the amplitude $C$ of the secondary eigenfunction to unity, since our final expressions do not depend on its value, and we omit the indices enumerating different secondary 
modes hereafter, assigning index $s$ to quantities belonging to the secondary mode where this matters. 

First, we calculate the difference $4\omega_p^2-\omega_j^2\equiv  4\omega_p^2-\omega_s^2$. Using equation (\ref{mp5}), we obtain
\begin{eqnarray}
4\omega_p^2-\omega_s^2={12 {\omega^0_p}^2\over z_r}\nu,
\label{mp12}
\end{eqnarray} 
where 
\begin{eqnarray*}
\nu&=&{3^{1/6}\over {2\sqrt{z_r}}}{\Gamma(1/3)\over \Gamma (2/3)}\left(\kappa (2k_p)-\kappa (k_p)\right){\left({\omega_p^0\over k_p}\right)}^{5/3}+
3{\omega^0_p}^2,
\end{eqnarray*}     
and where $\kappa (k)$ is given by (\ref{mp2}) and all quantities on the right hand side depend on $\omega^0_p$ and $k_p$. From (\ref{mp2}) we have $\kappa (2k_p)-\kappa (k_p)=k_p(1-e^{-2k_pz_c})/(1+e^{-2k_pz_c})$. This relation tells us that the correction proportional to $\lambda_*^2$ tends to zero 
when $z_c k_p \rightarrow 0$. In all other expressions below we can use $\omega^0_p$ and $k_p$, therefore, for
simplicity, we set $\omega\equiv \omega^0_p$ and $k \equiv k_p$ from now on. We substitute the expression for the source term (\ref{mp6a})
and (\ref{mp12}) in (\ref{mp11}) and remember that $\lambda_j$ there should be equal to $\omega_p^2$. We obtain
\begin{equation}
a_s={2 iz_r\over 3 \omega \nu }{\int^{z_r}_{0}{\bar v_p}^3dz\over \int^{z_r}_{0} z{\bar v_p}^2dz}C_p^{2}e^{2i\omega t}.
\label{mp13}
\end{equation} 
Note that since $\phi^{*}_s \propto e^{-2ik_px}$ and $S_{p}\propto e^{2ik_px}$ the integrands do not depend on $x$, so integration over 
$x$ is trivial. In order to evaluate the integrals entering (\ref{mp13}) we neglect $\omega^2$ in the expression for $y$ in terms of $z$ (see (\ref{b20})), and change the integration variable from $z$ to $y$. Since the integral in the numerator converges when $z_r/\lambda_* \rightarrow \infty$, its upper limit of integration can be extended to infinity. Numerical evaluation of $\int^{\infty}_{0}{\bar v_p}^3dy$ shows that its value is close to 0.2. On the other hand the integral in the denominator diverges as 
$(z_r/\lambda_*)^{3/2}$ in the same limit. Therefore, we use the limiting value of the ratio $\int^{z_r/\lambda_*}_{0}y{\bar v_p}^2/(z_r/\lambda_*)^{3/2}dy$, which is close to $0.1$ to obtain  ${\int^{z_r}_{0}{\bar v_p}^3dz\over \int^{z_r}_{0} z{\bar v_p}^2dz}\approx 2{\lambda_*^{1/2}\over z_r^{3/2}}$.
Substituting this in (\ref{mp13}) we get
\begin{equation}
a_s\approx {4i\over 3}{1 \over z_r^{1/2}k^{1/3}\omega^{2/3}\nu} C_p^2e^{2i\omega t}.
\label{mp14}
\end{equation}      
We expect nonlinear behaviour to be important when $|a_s| \gtrsim C_p$, and potentially even for amplitudes quite a bit smaller than this. Equation (\ref{mp14}) tells that the wave amplitude $C_p$ in this case should exceed its critical value 
\begin{equation}
C_{p}^{crit}={3\over 4}z_r^{1/2}k^{1/3}\omega_p^{2/3}\nu,
\label{mp15}
\end{equation} 
for these nonlinearities to be important. Above this amplitude, the secondary waves have amplitudes that are comparable with or exceed the primary wave amplitude.

\section{Relation to the tidal problem and criterion for the transition to nonlinearity}
\label{relation}

\subsection{Relation of our local model results to a spherical star}
\label{local_spherical} 
In the previous section we used dimensionless units in which spatial scales were expressed in terms of the radius of the base of the convection zone, $r_c$, and temporal ones were expressed in terms of the characteristic frequency $\omega_{c}=\sqrt{A_cr_c}$, where $A_c$ was defined in eq. (\ref{e14}). Bearing this in mind, we can relate results obtained from the local model in the two previous sections to the global normal modes of a sun-like star. For that we compare (\ref{b20}) with equation (91), and (\ref{mp5}) with equations (103-106) of IPCh, respectively. From the comparison of the first pair of equations, we see that they coincide when the term proportional to
$\omega^2$ is neglected in the argument of the Airy functions in (\ref{b20}) and we set $k=\sqrt{\Lambda}$, where $\Lambda $ is defined in IPCh as an eigenvalue of Laplace's tidal equation, and it can be expressed as $\Lambda=l(l+1)$, where $l$ is the spherical harmonic degree for a non-rotating star. Note that the term $\propto \omega^2$ was neglected in IPCh because of its smallness, but it is, however, important for our purposes, since it gives a potentially important correction in the expression (\ref{mp15}) for the quantity $\nu$.  
Comparing the expression for $\omega_n^0$ with eq. (103) of IPCh, we see that they are equivalent to each
other provided we make the following redefinitions in (\ref{mp5}): $n \rightarrow n+2$ and $z_r \rightarrow {\left({3\over 2}{I\over \omega_c}\right)}^{2/3}$, where $I$ is defined in (104) of IPCh. Comparing eq. (105) of IPCh and the term proportional to the factor $\kappa$ in the expression for $\nu$ in (\ref{mp15}), these equations formally lead to the same frequency correction when $\kappa=r_cB_c$, where the quantity $B_c$ defined in equation (100) of IPCh. Note, that the numerical values are, of course, different, since $B_c$ was calculated in spherical geometry and for a particular size of convective envelope.

\begin{figure}
\includegraphics[width=0.5\textwidth]{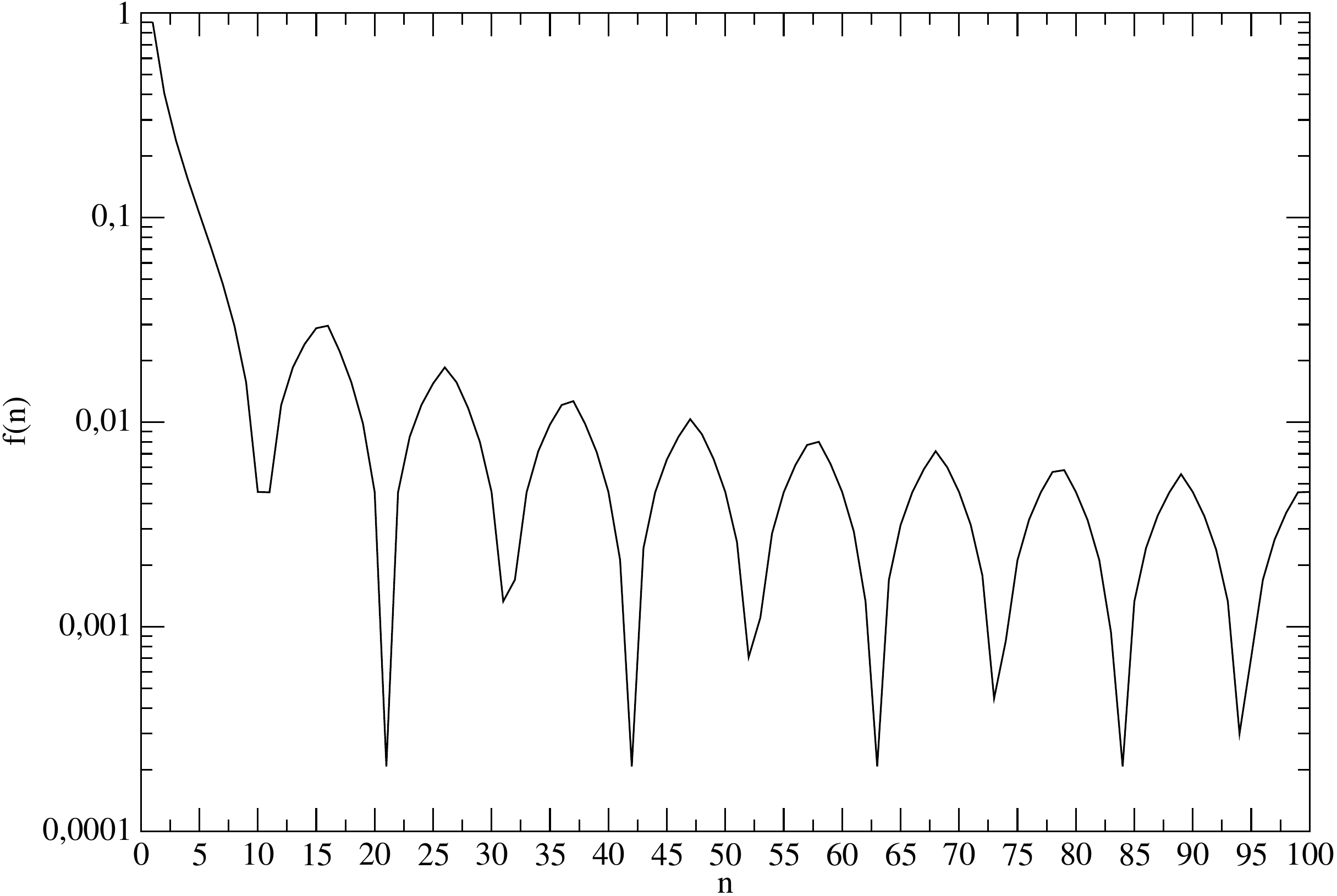}
\caption{The function $f(n)$ defined in eq. (\ref{mpn4}) is shown as a function of $n$.}
\label{fn}
\end{figure}

It is very important to stress that the planar geometry used in the previous sections can be shown to be fully equivalent to
the spherical one only in the limit $l \rightarrow \infty$. For the tidal problem considered in this paper, however,
$l=2$ for the primary mode. On the other hand, for the secondary mode $l$ should be equal to $4$, which results 
in an additional correction in the expression for the quantity $\nu$ characterising the frequency difference between 
primary and secondary modes. In order to calculate it, we start from the expression for the eigenfrequencies $\omega_{0,n}$
calculated in IPCh in the WKBJ approximation, and given in their equation (103): 
\begin{equation}
\omega_{0,n}={\sqrt{l(l+1)}I\over \pi (n+(l+1)/2+1/12)}.
\label{mpn1}
\end{equation} 
Note that for simplicity we neglect their correction $\delta \phi(\omega_{0,n})$, since it is already taken into account
in the expression for $\nu$ discussed above. We assume $n\gg 1$ for the validity of the WKBJ approximation, therefore the terms in the denominator can be considered as small corrections. We can discard them altogether if we note that when $l=2$ and $4$ for the primary and secondary modes, they differ by one and this difference can be absorbed into a redefinition of the mode number $n$. The rest don't
give any contribution at leading order in the small parameter $\Delta_l=\sqrt{{6\over 5}}-1\sim 0.095$ arising from 
the difference between the values $2\sqrt{l(l+1)}$ and $\sqrt{l(l+1)}$ in the numerator in (\ref{mpn1}) for $l=2$ and $4$, respectively. Taking this into account we can use
\begin{equation}
\omega_{0,n}={\sqrt{l(l+1)}I \over \pi n}.
\label{mpn2}
\end{equation}      
when calculating the frequency difference $2\omega_p-\omega_s$ and, accordingly, the additional correction to $\nu$
setting $l=2$ and $4$ for $\omega_p$ and $\omega_s$, respectively. For the primary mode we set $n=n_p$
and for the secondary mode $n=n_s \equiv n_p-k$, where it is assumed that $k\ll n_p$, and $k$ should be chosen in such a way
that the absolute value of $2\omega_p-\omega_s$ is minimised. 
Taking into account only the leading terms in $\Delta_{l}$, we obtain
\begin{equation}
2\omega_p-\omega_s={2 \sqrt 6 I\over \pi n_p}f(n_p)=2\omega_pf(n_p),
\label{mpn3}
\end{equation} 
where
\begin{equation}
f(n)=\textrm{min}_k\left\vert\sqrt{{6\over 5}}-1-\frac{k}{n}\right\vert.
\label{mpn4}
\end{equation} 
We show the dependency of $f(n)$ on $n$ in Fig. \ref{fn}. 
Comparing (\ref{mpn3}) with the expression (\ref{mp12}) for 
$\nu$ and remembering that $z_c ={({3\over 2}{I\over \omega_c})}^{2/3}$, we see that $\nu$ should contain the
additional term 
\begin{equation}
{\left({2\over 3}\right)}^{1/3}{\left({I\over \omega_c}\right)}^{2/3}f(n_p).
\label{mpn5}
\end{equation}
As we mentioned above for the tidal problem $\omega_p \approx 2\Omega_{orb}$. Using (\ref{mpn2}) with $l=2$ 
we can express $n_p$ in terms of $\Omega_{orb}$ and consider $f(n)$ to be a function of the orbital frequency. We have
\begin{equation}
n_p=\left[{\sqrt{6} I \over 2\pi \Omega_{orb}}\right],
\label{mpn6}
\end{equation}
where $[.]$ implies that only the integer part of the expression is used. 

Restoring physical units, using the redefinitions mentioned above, and substituting 
(\ref{mpn5}) in (\ref{mp15}), we obtain from (\ref{mp12}) and (\ref{mp15}):    
\begin{eqnarray}
\nonumber
\nu&=&{3^{\frac{1}{6}}\Gamma (\frac{1}{3})\over 2 \cdot6^{\frac{5}{6}}\Gamma \left(\frac{2}{3}\right)}{\left({{2\over 3}{\omega_c\over I}}\right)}^{\frac{1}{3}} B{\left({\omega_p\over \omega_c}\right)}^{\frac{5}{3}} +3{\left({\omega_p\over \omega_c}\right)}^2\\
&&\hspace{1cm}+{\left({2\over 3}\right)}^{\frac{1}{3}}{\left({I\over \omega_c}\right)}^{\frac{2}{3}}f(\Omega_{orb}),
\label{mp16}
\end{eqnarray} 
where $B=r_c(B_c(2\cdot 6^{\frac{1}{2}})-B_c(6^{\frac{1}{2}}))$, 
and
\begin{equation}
C_{p}^{crit}=3^{\frac{3}{2}}2^{-\frac{13}{6}}{\left({\omega_p\over \omega_c}\right)}^{\frac{2}{3}}{\left({I\over \omega_c}\right)}^{\frac{1}{3}}\nu \omega_cr_c.
\label{mp17}
\end{equation} 

\subsection{The criterion for transition to nonlinearity for a single secondary mode}

The above expression (\ref{mp17}) should be compared with the amplitude of the radial component of the linear tidal response. This amplitude, $U^z=i\omega_p
{\bmth \xi}^r$
can be calculated using (\ref{e2nn}), where $\omega_p=2\Omega_{orb}$ and 
we should take the limit $r\rightarrow r_c$ when evaluating $\xi$.  The corresponding expression is given by equation (101) of IPCh: 
\begin{equation}
\xi \approx {(-1)^n\over \Gamma ({2\over 3})}\sqrt{\pi }{\left({3\omega_p\over \omega_c}\right)}^{-\frac{1}{6}}
{6^{\frac{1}{12}}\over \omega_c^{\frac{1}{2}}\rho_c^{\frac{1}{2}}r_c^{\frac{3}{2}}}C_{WKBJ}.
\label{e15}
\end{equation}
Using equation (\ref{e10}) for $C_{WKBJ}$ we obtain from (\ref{e15})
\begin{equation}
\xi \approx {(-1)^n\over \Gamma ({2\over 3})}3^{-\frac{1}{12}}2^{\frac{7}{12}}\sqrt{\pi}{\left({\omega_p\over \omega_c}\right)}^{\frac{5}{6}}{\left({\omega_c\over I}\right)}^{\frac{1}{2}}{n^{\frac{1}{2}}\over \rho^{\frac{1}{2}}_c r_c^{\frac{3}{2}}}.
\label{mp18}
\end{equation}
We substitute (\ref{mp18}) in (\ref{e2nn}), use $\omega_p=2\Omega_{orb}$, and multiply the result by $2\Omega_{orb}$, to obtain
\begin{eqnarray}
\nonumber
U^z&=&2i\Omega_{orb}{\bmth \xi}^r(r_c) \\
\nonumber
&=& i{(-1)^n\over \Gamma ({2\over 3})}3^{\frac{11}{12}}2^{-\frac{19}{12}}\pi^{\frac{3}{2}}\left({q\over 1+q}\right)
 \\
&& \hspace{0.6cm}
 \times \left({\omega_c\over 
|d\omega/dj|}\right){\left({\omega_c\over I}\right)}^{\frac{1}{2}}{\left({\Omega_{orb}\over \omega_c}\right)}^{\frac{17}{6}}\hat Q {\omega_c\over \rho_c^{\frac{1}{2}}r_c^{\frac{3}{2}}}.
\label{mp19}
\end{eqnarray}
Note that since $\hat Q$ has dimensions $g^{1/2}\cdot \mathrm{cm}$, equation (\ref{mp19}) has the correct dimensions $\mathrm{cm/s}$. When (\ref{mp19}) is larger than than (\ref{mp17}) we assume that a fully non-linear regime sets in. From the condition $|U^{z}| > C_{p}^{crit}$ we have
\begin{eqnarray}
\nonumber
{q\over 1+q} > C_{crit, c}&=&3^{\frac{7}{12}}2^{\frac{1}{12}}\pi^{-\frac{3}{2}}\Gamma\left(\frac{2}{3}\right){|d\omega/dj|\over \omega_c} \\ &&\times {\left({I\over \omega_c}\right)}^{\frac{5}{6}}{\left({\Omega_{orb}\over \omega_c}\right)}^{-\frac{13}{6}} \nu {\rho_c^{\frac{1}{2}}r_{c}^{\frac{5}{2}}\over \hat Q}.  
\label{mp21}
\end{eqnarray}
Finally, we use (\ref{e11}) to obtain
\begin{equation}
 C_{crit, c}=3^{\frac{1}{12}}2^{\frac{19}{12}}\pi^{-\frac{1}{2}}\Gamma\left(\frac{2}{3}\right){\left({I\over \omega_c}\right)}^{-\frac{1}{6}}{\left({\Omega_{orb}\over \omega_c}\right)}^{-\frac{1}{6}} \nu {\rho_c^{\frac{1}{2}}r_{c}^{\frac{5}{2}}\over \hat Q}.  
\label{mp22}
\end{equation}
This is an amplitude criterion for our secondary super-harmonic wave to have a comparable (or larger) amplitude to the primary tidal wave, above which we expect consideration of nonlinear effects to be essential.

\subsection{The criterion for transition to nonlinearity with a dense spectrum of secondary modes}

If the spectrum of eigenmodes is sufficiently dense, the primary mode can effectively excite a number of neighbouring secondary modes instead of just the one having a frequency that most closely matches the primary frequency. As discussed in Section \ref{dense}, in order for the spectrum to be sufficiently dense, either the ratio $\delta$ of the value of the frequency offset (proportional to the quantity $\nu$ defined in (\ref{mp12})) to the distance between neighbouring eigenfrequencies, $d\omega_j/dj$, is large, or the inverse decay time, $\omega_{\nu}$, is larger than $d\omega_j/dj$. While the former condition can be straightforwardly calculated within the framework of our model, the latter condition assumes that the wave packet composed of secondary waves with approximately the same frequency decays during its travel across a star. Since this has been assumed for the primary wave, it is reasonable to make the same assumption for these secondary waves as well. In both cases, the modified values of $C_{crit, c}$ are supposed to be smaller than the values in (\ref{mp22}), so we can consider (\ref{mp22}) as a conservative estimate of $C_{crit}$. On the other hand, our estimate here can be considered a more `optimistic' estimate, suggesting what can in principle be achieved from the effect discussed in this paper.   

The quantity $\delta$ defined below eq. (\ref{l6}) can be calculated using (\ref{e11}) and (\ref{mp12}).
From equation (\ref{mp12}), it follows that the distance between the doubled primary and secondary frequencies, $\Delta \omega_s=2\omega_p-\omega_s$, can be expressed as
\begin{equation}
\Delta \omega_s=3^{\frac{1}{3}}2^{\frac{5}{3}}{\left({\omega_c\over I}\right)}^{\frac{2}{3}}\nu \Omega_{orb},  
\label{mp23}
\end{equation}
where we remember that $z_{r}={({3\over 2}{I\over \omega_c})}^{\frac{2}{3}}$ and $\omega_p=2\Omega_{orb}$. Now we note that 
there should be $2\sqrt{6}$ instead of $\sqrt{6}$ in the denominator of (\ref{e11}), see also eq. (125) of IPCh, and we
divide (\ref{mp23}) by this expression to obtain  
\begin{equation}
 \delta={2^{\frac{7}{6}}\cdot 3^{\frac{5}{6}}\over \pi}{\left({I\over \omega_c}\right)}^{\frac{1}{3}}\nu {\omega_c\over \Omega_{orb}}.  
\label{mp24}
\end{equation}
   
From our discussion in Section \ref{dense}, it follows that when $\delta $ and/or $\kappa $ are large $\Delta \omega_s$ in the expressions
for the amplitude of excited modes, as in e.g. (\ref{mp13}), should be substituted by $({d\omega_s/dj})(S(\delta, \kappa))^{-1}$,
where $S(\delta, \kappa)$ is given by (\ref{l7}) and we remember that $d\omega_s/dj={2\pi\Omega^2_{orb}\over \sqrt{6}I}$ for secondary modes. Since $\nu$ is proportional to $\Delta \omega_s$ it should be changed accordingly in all expressions,  including in (\ref{mp22}). Therefore, in order to account for the excitation of many near-resonant secondary modes
we can express $\nu $ in terms of $\Delta \omega_s$ using (\ref{mp23}), and make the substitution
\begin{equation}
 \nu \rightarrow {\pi \over 3^{\frac{5}{6}}\cdot 2^{\frac{7}{6}}}|S|^{-1}{\Omega_{orb}\over I^{\frac{1}{3}}\omega_c^{\frac{2}{3}}}  
\label{mp25}
\end{equation} 
in (\ref{mp22}), where we use the absolute value of $S$, since it can be a complex quantity. In this way, we obtain
\begin{equation}
C^{dense}_{crit, c}=3^{-\frac{3}{4}}2^{\frac{5}{12}}\pi^{\frac{1}{2}}\Gamma \left({2\over 3}\right){\left({I\over \omega_c}\right)}^{-\frac{1}{2}}|S|^{-1}{\left({\Omega_{orb}\over \omega_c}\right)}^{\frac{5}{6}}{\rho_c^{\frac{1}{2}}r_{c}^{\frac{5}{2}}\over
\hat Q}.  
\label{mp26}
\end{equation}
It is very important to stress here, that $C^{dense}_{crit}$ given by (\ref{mp26}) should be considered as a qualitative indication of the importance of having a dense spectrum of secondary waves rather than a quantitative criterion, since the derivation of this expression is based on a number of simplifying assumptions. In particular, we assume that the summation of the series in (\ref{l6}) can be formally extended to infinity.

\section{Numerical study of weakly nonlinear excitation of super-harmonic secondary waves}
\label{numerical}

In this section we briefly present some numerical calculations to verify the analytical results obtained in \S~\ref{boussi}--\ref{weaklynonlin}. We first solve numerically the generalised eigenvalue problem for time-harmonic solutions proportional to $\mathrm{e}^{i (\omega t + k_p x)}$ to the linearised system of equations
\begin{eqnarray}
\label{mom1}
i\omega {\bf U}&=&-\nabla P + b{\bf e}_z-\gamma {\bf U}, \\
i\omega b&=&-N^2(z) U^{z}, \\
 \quad \nabla \cdot {\bf U}&=&0,
\end{eqnarray}
where $\omega\in \mathbb{C}$ is the eigenvalue,
\begin{eqnarray}
N^2(z)=\begin{cases}
z, \quad \;\;0\leq z\leq z_r, \\
0, \quad -z_c \leq z<0,
\end{cases}
\end{eqnarray}
subject to $U^z(z=-z_c)=U^z(z=z_r)=0$. We have introduced a frictional damping term $-\gamma {\bf U}$ where $\gamma=10^{-4}$ is a constant (analogous to $\omega_\nu$ in section 2, which is simpler numerically than including viscosity or thermal diffusion). This problem is discretised in $z$ using a Chebyshev collocation method with $\mathcal{N}+1$ points \citep{Boyd2001}, with $\mathcal{N}=200$, and we choose $k_p\equiv k =z_c=z_r=1$ to define our choice of space and time units (noting that we then have $\mathrm{max}[N(z)]=1$). The solution of this eigenvalue problem gives us a set of eigenvalues $\{\omega_i\}$ with corresponding eigenvectors $\{({\bf U}, P, b)_i\}$. We order the eigenvalues $\omega_i\leq \mathrm{max}[N(z)]$ in descending order using their real part since the modes with the largest frequencies are those that are best resolved numerically. 

We then choose a single mode $i$ as our ``primary mode", set $\omega_p=\mathrm{Re}[\omega_i]$, and substitute the corresponding eigenfunction $\bar{v}_p(z)=U^z_i$ (without loss of generality, this is normalised so that it is real with a maximum magnitude that matches (\ref{b20})) into the right hand side of equation (\ref{mp6a}), to which we also add the additional frictional damping term $\gamma \Delta \dot{U}^z$ on the left hand side (equivalent to the term $-\gamma{\bf U}$ in \ref{mom1}). We solve this equation for the ``secondary modes" as an initial value problem with boundary conditions $U^z(z=-z_c,t)=U^z(z=z_r,t)=0$, initialising our solution with $U^z(z,t=0)=0$ so that the forcing excites all of the modes until they are damped by friction. To do this, we use Chebyshev collocation in $z$ with $\mathcal{N}+1$ points (again with $\mathcal{N}=200$) and a second-order central difference scheme for time integration. This equation is integrated until a time $t=5000$ and the frequency power spectrum ($|\hat{U}^z|^2$) is computed using a discrete Fourier transform of the signal at the location $z=0.1$.

\begin{figure}
  \begin{center} 
      \subfigure{\includegraphics[trim=5cm 0cm 5cm 0cm, clip=true,width=0.49\textwidth]{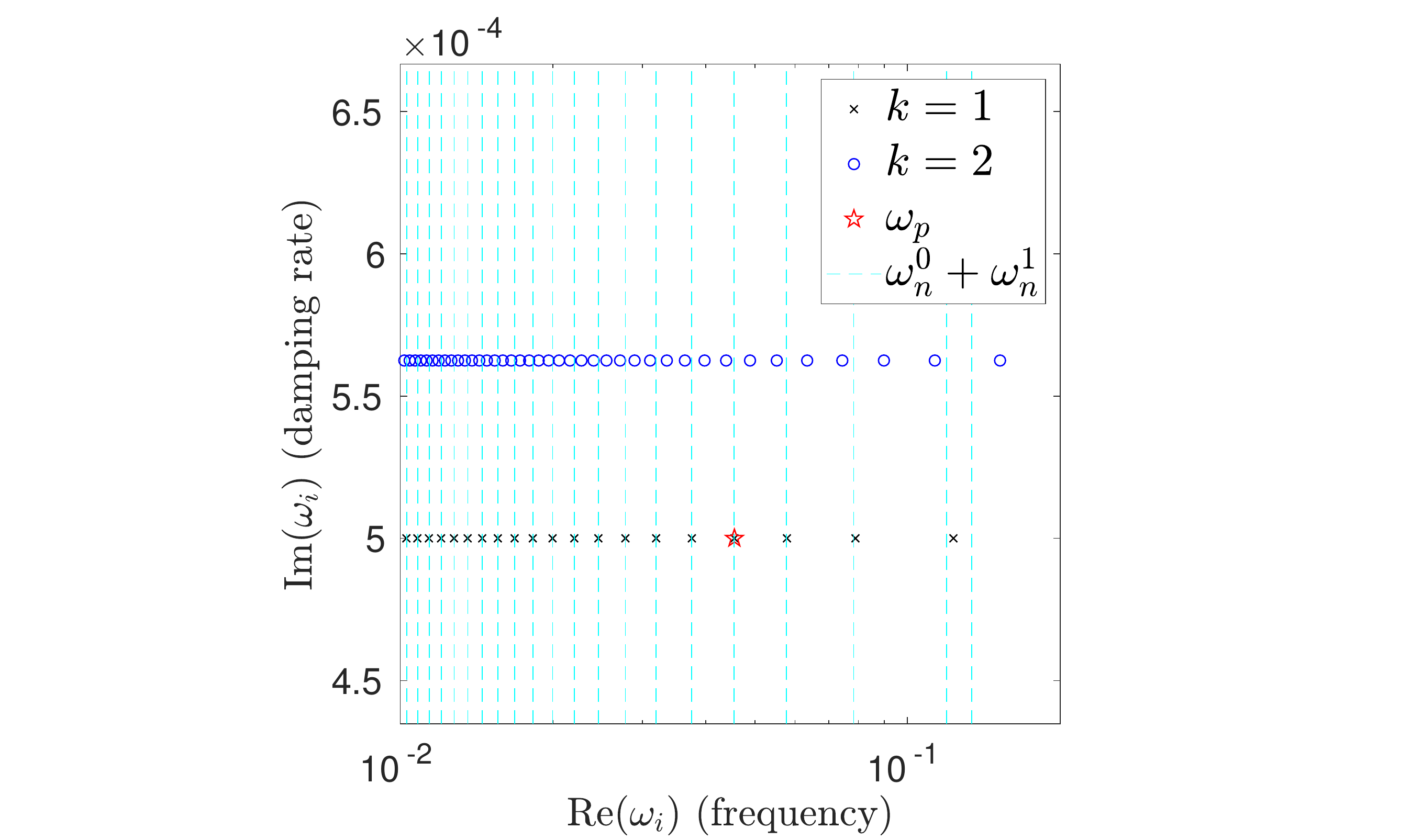}}
       \end{center}
  \caption{Mode spectrum showing the real and imaginary parts of the eigenfrequencies $\omega_i$ for $k=1$ (black crosses) and $k=2$ (blue circles) with $z_c=z_r=1, \gamma=10^{-4}, \mathcal{N}=200$. The cyan dashed lines are the analytical predictions for $k=1$ using \ref{mp5}, which agree very well for low frequencies. The red star highlights the mode with $\omega_p=\mathrm{Re}[\omega_i]\approx 0.0456$.}
  \label{eval}
\end{figure}

\begin{figure*}
  \begin{center} 
      \subfigure[Primary $U^z$ profile]{\includegraphics[trim=2cm 0cm 4cm 0cm, clip=true,width=0.49\textwidth]{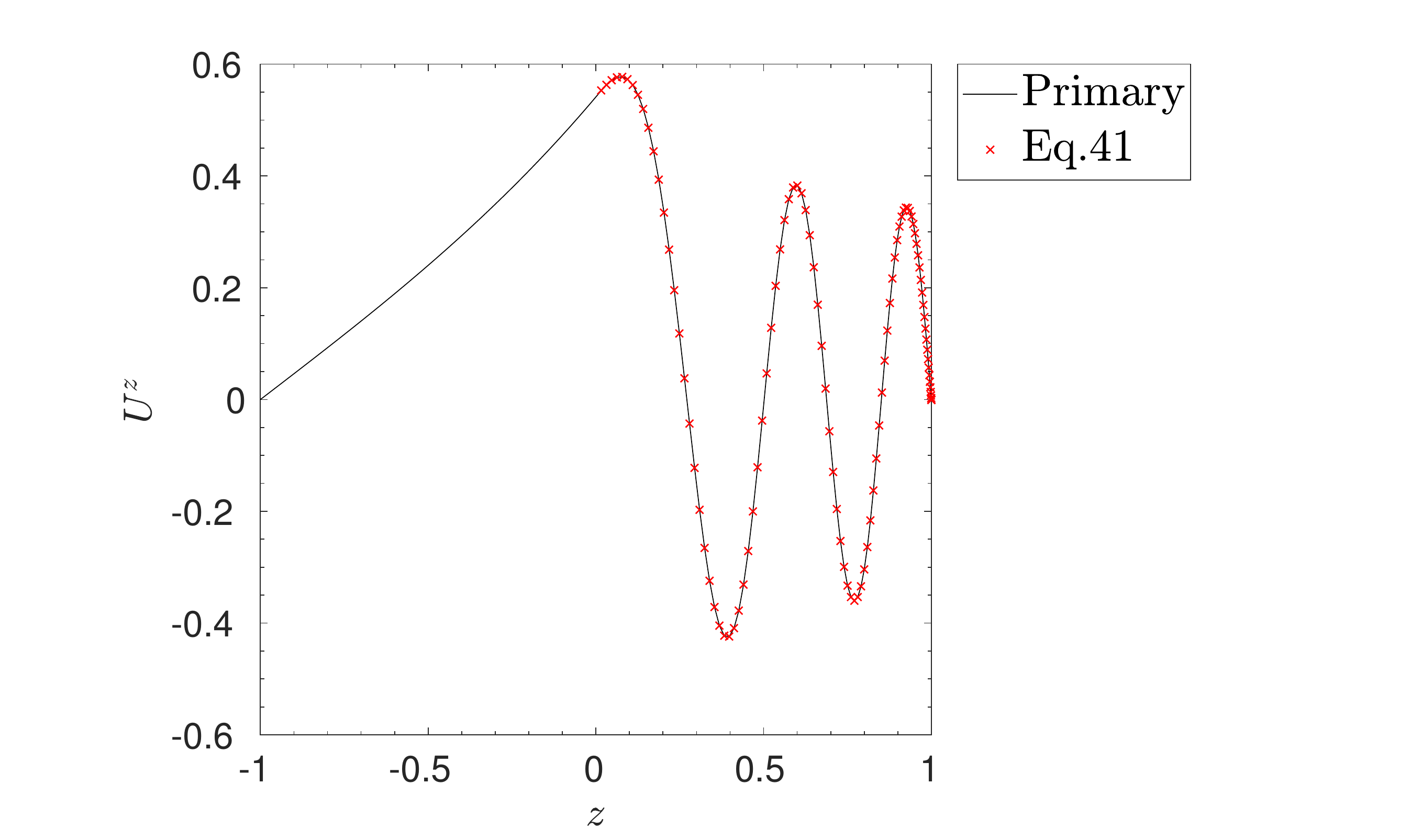}\label{2a}}
      \subfigure[Secondary $U^z$ vs $z$ at $t=5000$.]{\includegraphics[trim=4cm 0cm 4cm 0cm, clip=true,width=0.49\textwidth]{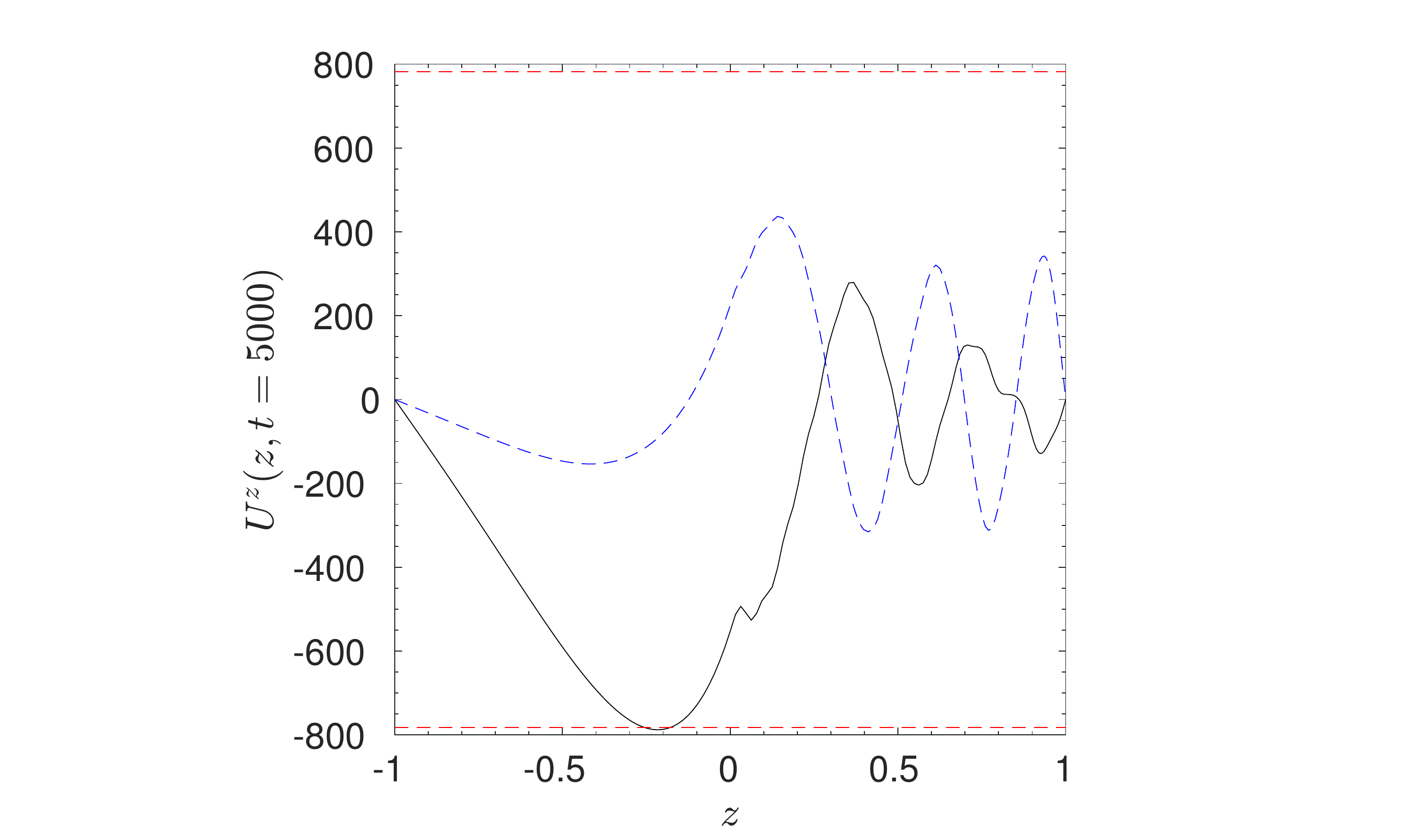}\label{2b}}
      \subfigure[Secondary $U^z$ vs $t$ at $z=0.1$.]{\includegraphics[trim=4cm 0cm 5cm 0cm, clip=true,width=0.49\textwidth]{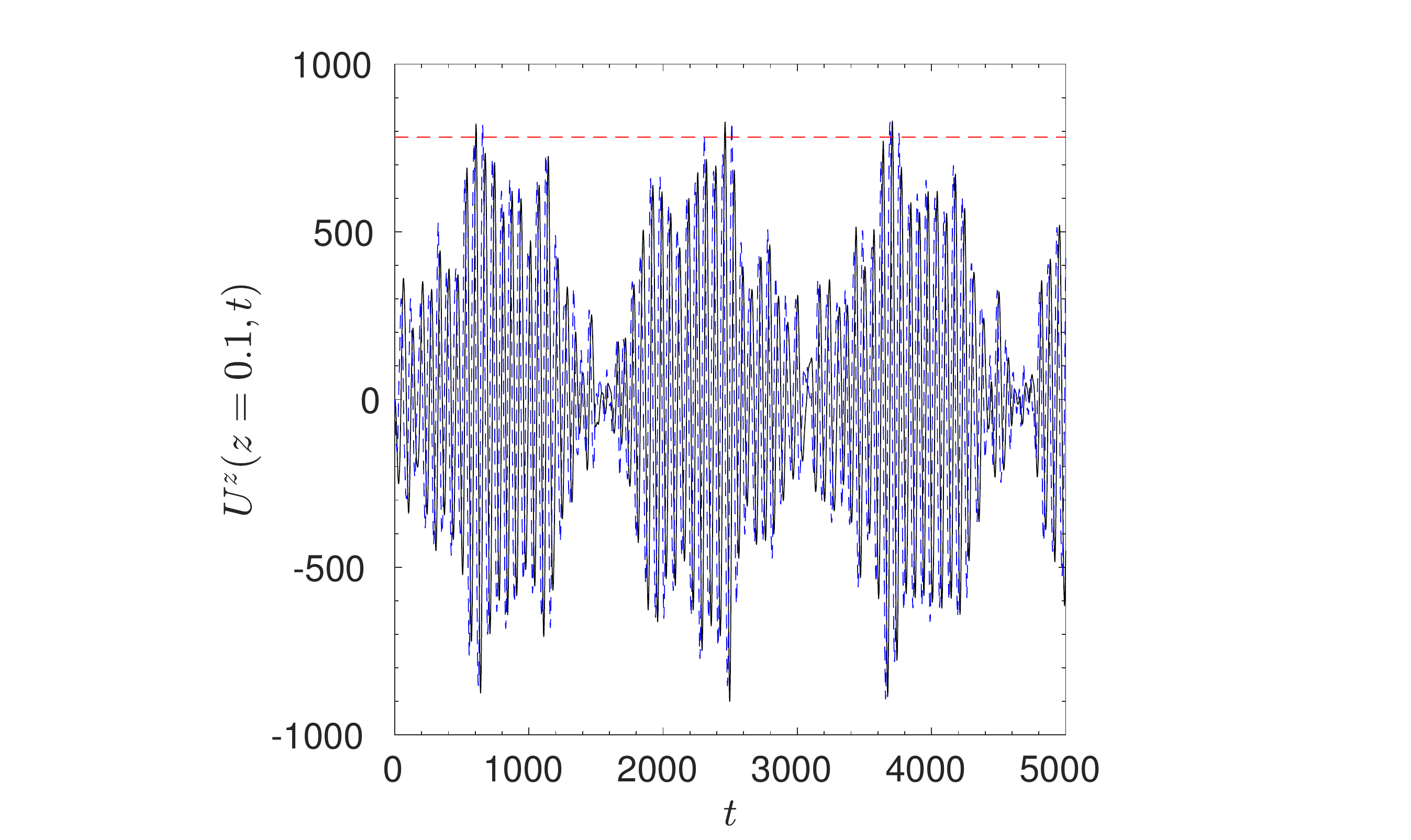}\label{2c}}
      \subfigure[Frequency power spectrum of $U^z(z=0.1,t)$.]{\includegraphics[trim=3cm 0cm 4cm 0cm, clip=true,width=0.49\textwidth]{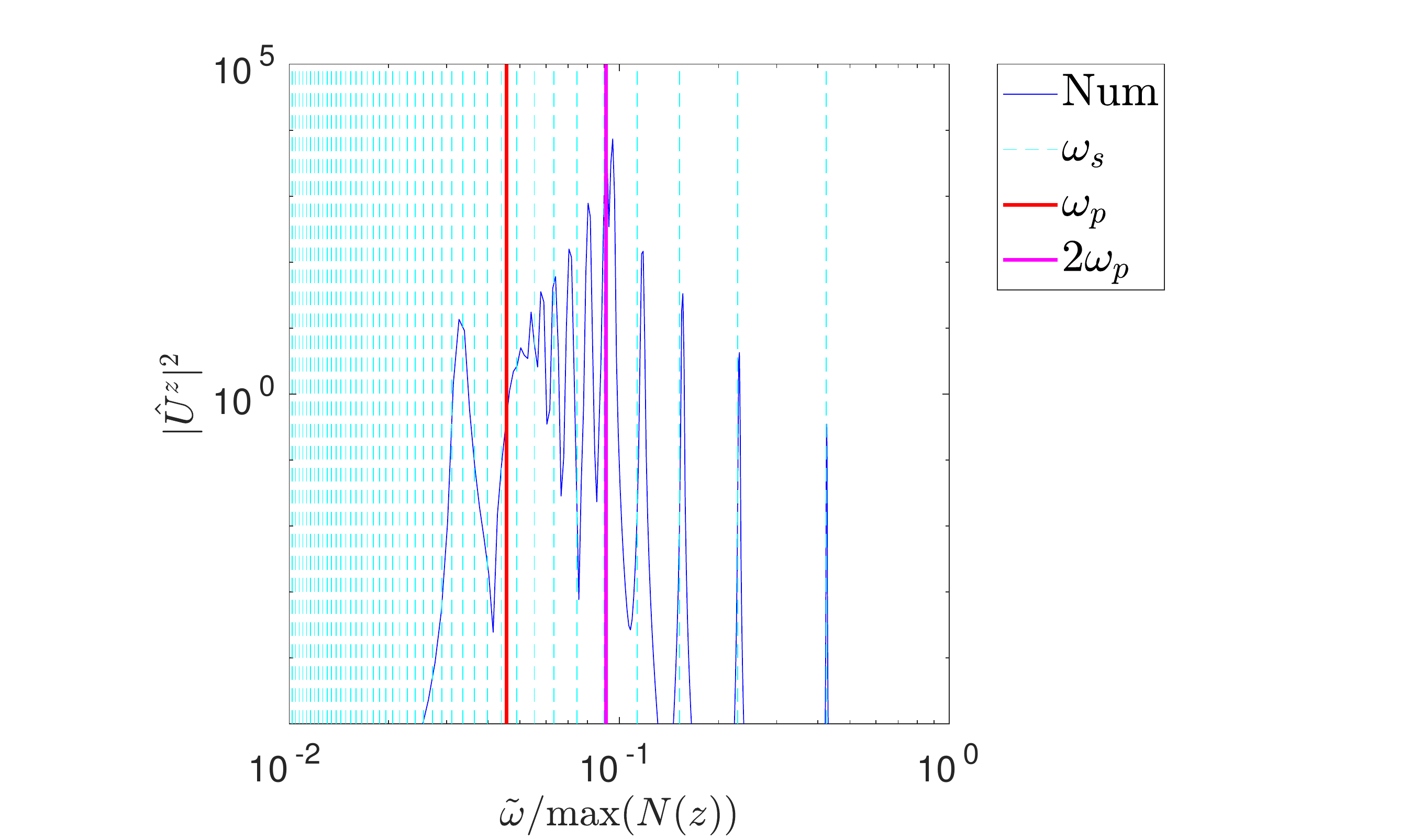}\label{2d}}
    \end{center}
  \caption{(a): Comparison of (normalised) primary mode eigenfunction $U^z$ with the analytical prediction in (\ref{b20}) for the mode with $\omega_p=\mathrm{Re}[\omega_i]\approx 0.0456$. (b): Real (solid black line) and imaginary (dashed blue line) parts of secondary mode $U^z$ as a function of $z$ at $t=5000$, compared with the prediction (red dashed line) for $a_s$ from (\ref{mp14}). (c): Real (solid black line) and imaginary (dashed blue line) parts of secondary mode $U^z$ as a function of $t$ at $z=0.1$ in the stable layer. (d): Frequency power spectrum of $U^z$, $|\hat{U}^z|^2$, at $z=0.1$ as a function of angular frequency $\tilde{\omega}$. Otherwise same parameters as Fig.~\ref{eval}.}
  \label{Fig2}
\end{figure*}

\begin{figure*}
  \begin{center} 
      \subfigure[Primary $U^z$ profile]{\includegraphics[trim=2cm 0cm 4cm 0cm, clip=true,width=0.49\textwidth]{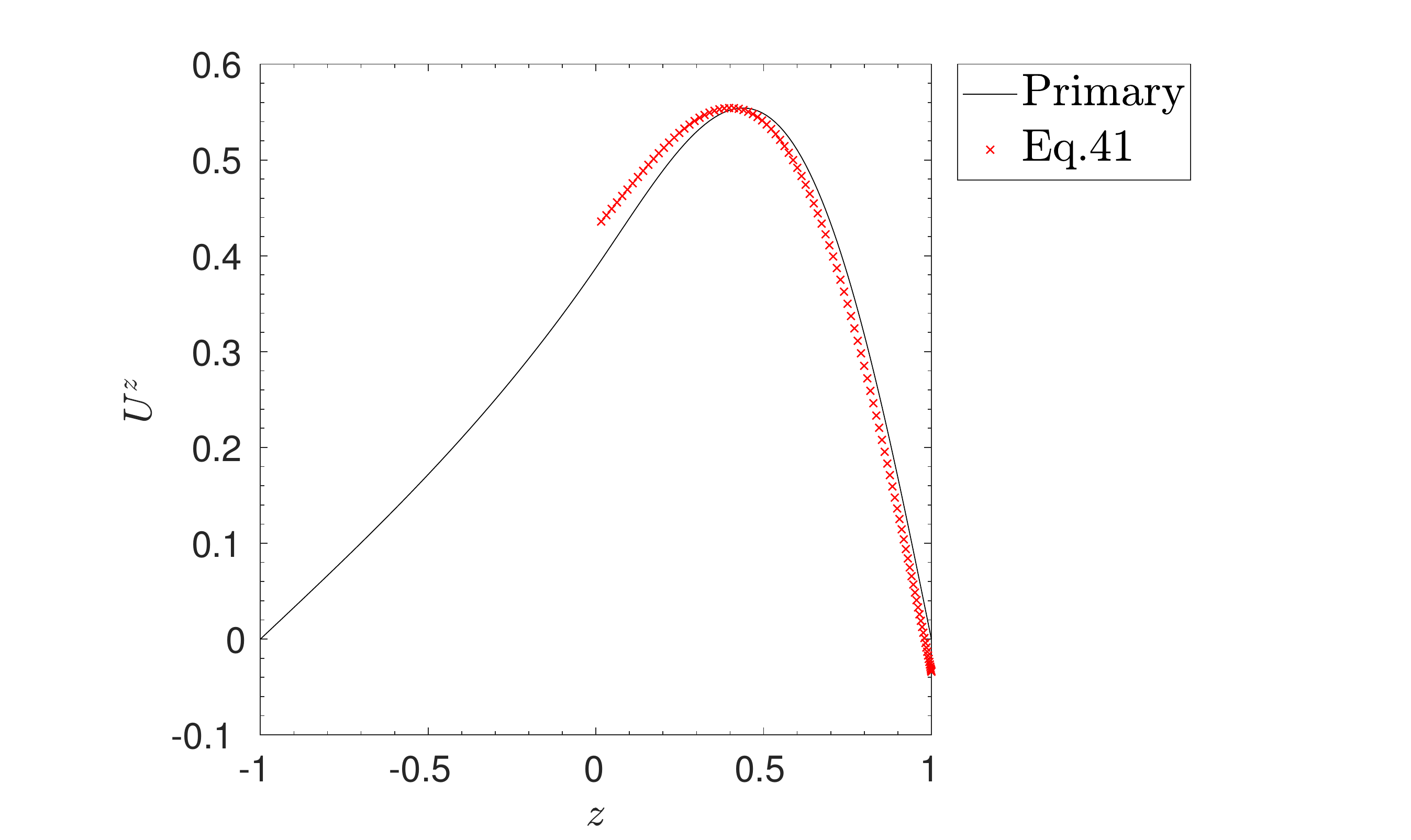}\label{3a}}
      \subfigure[Secondary $U^z$ vs $z$ at $t=5000$.]{\includegraphics[trim=4cm 0cm 4cm 0cm, clip=true,width=0.49\textwidth]{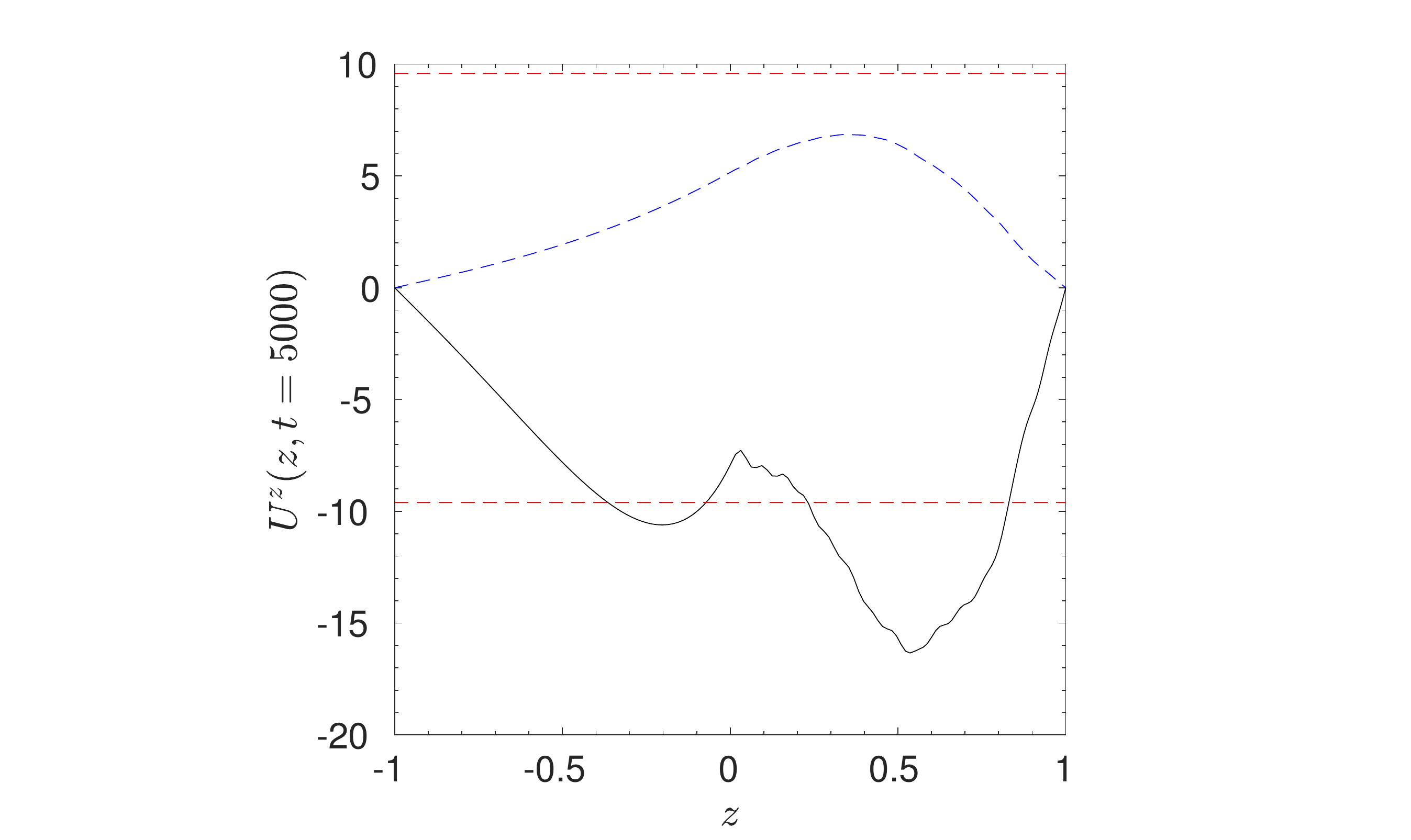}\label{3b}}
      \subfigure[Secondary $U^z$ vs $t$ at $z=0.1$.]{\includegraphics[trim=4cm 0cm 5cm 0cm, clip=true,width=0.49\textwidth]{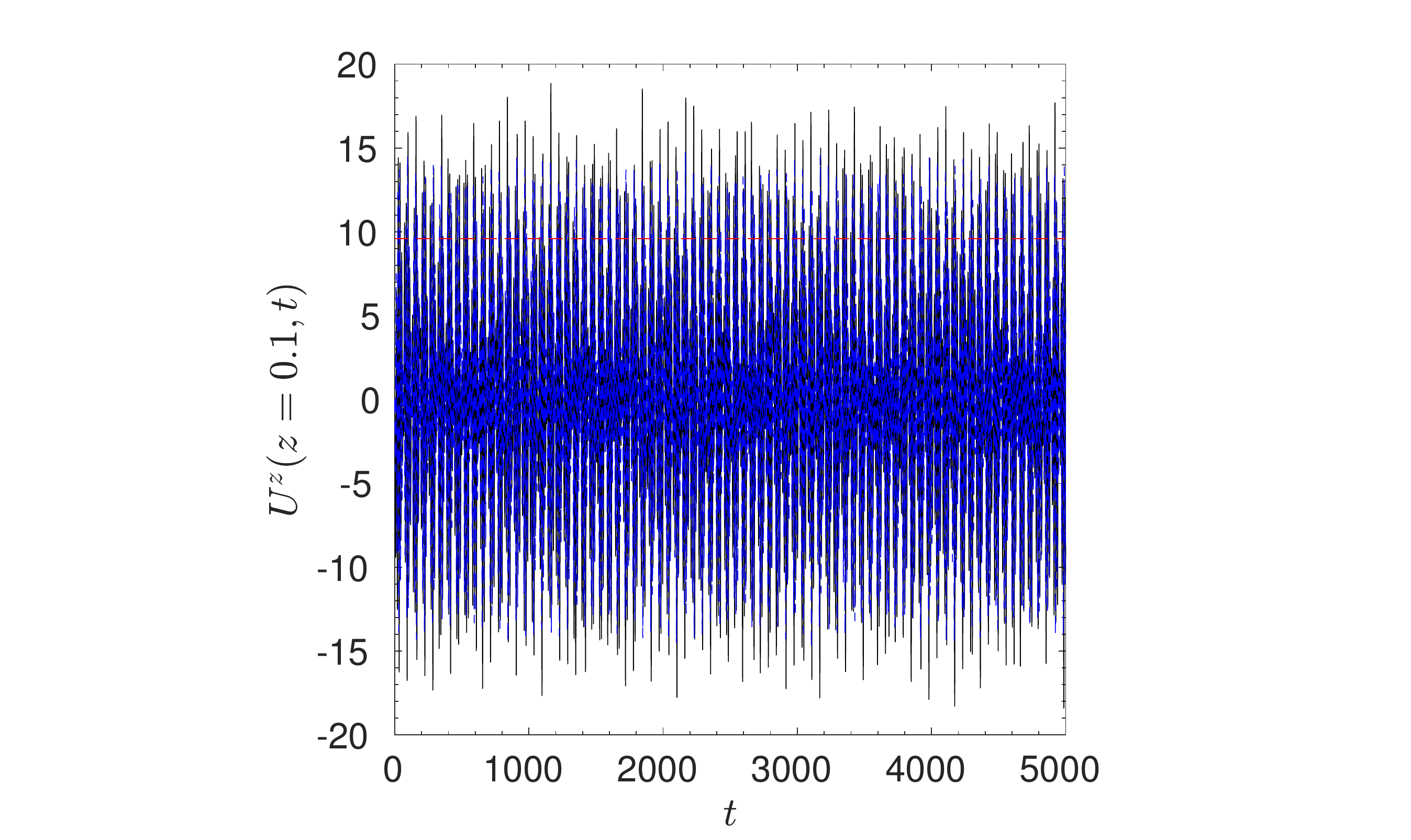}\label{3c}}
      \subfigure[Frequency power spectrum of $U^z(z=0.1,t)$.]{\includegraphics[trim=3cm 0cm 4cm 0cm, clip=true,width=0.49\textwidth]{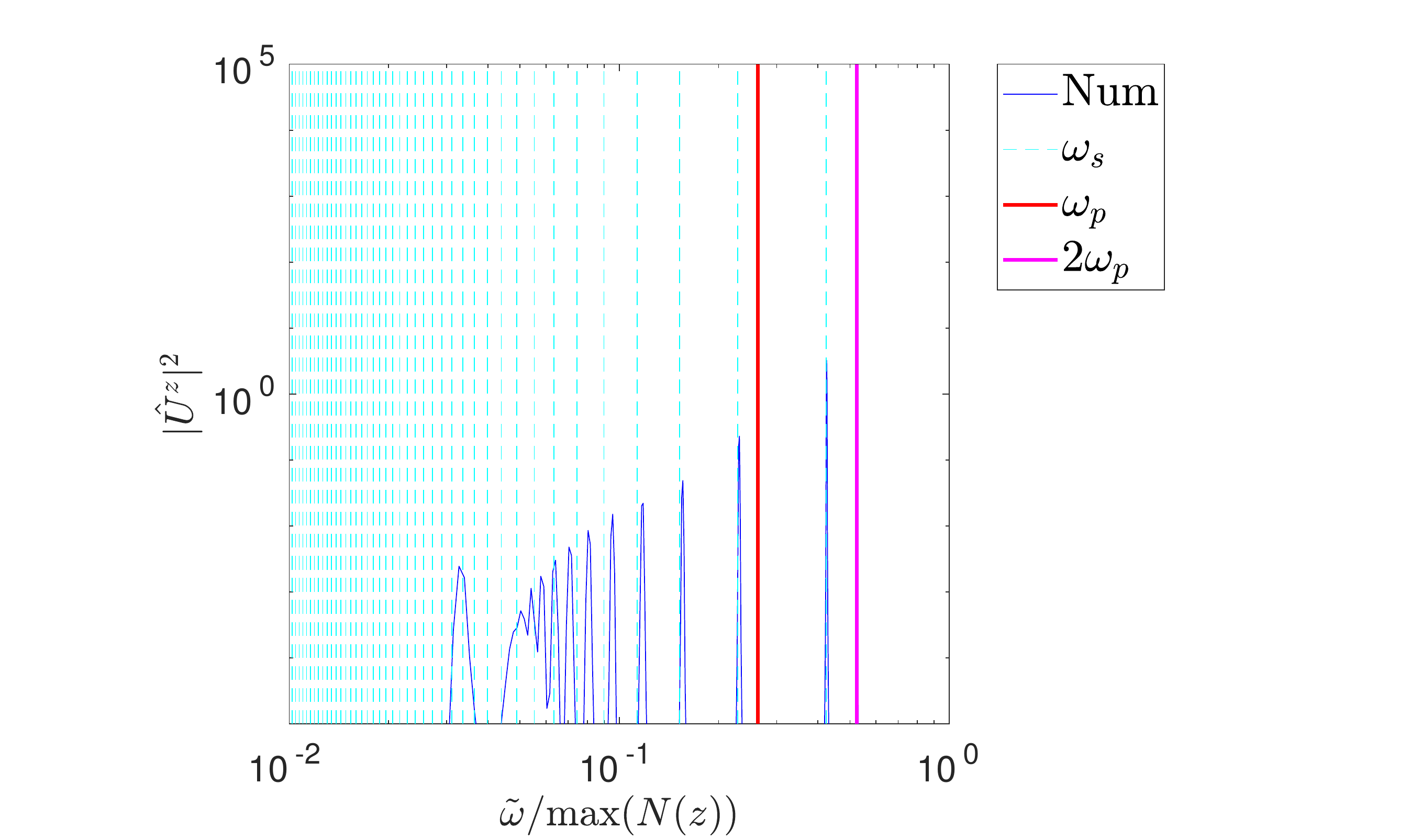}\label{3d}}
    \end{center}
  \caption{Same as Fig.~\ref{Fig2} but for a mode with $\omega_p=\mathrm{Re}[\omega_i]\approx0.263$.}
  \label{Fig3}
\end{figure*}

\begin{figure*}
  \begin{center} 
      \subfigure[Primary $U^z$ profile]{\includegraphics[trim=2cm 0cm 4cm 0cm, clip=true,width=0.49\textwidth]{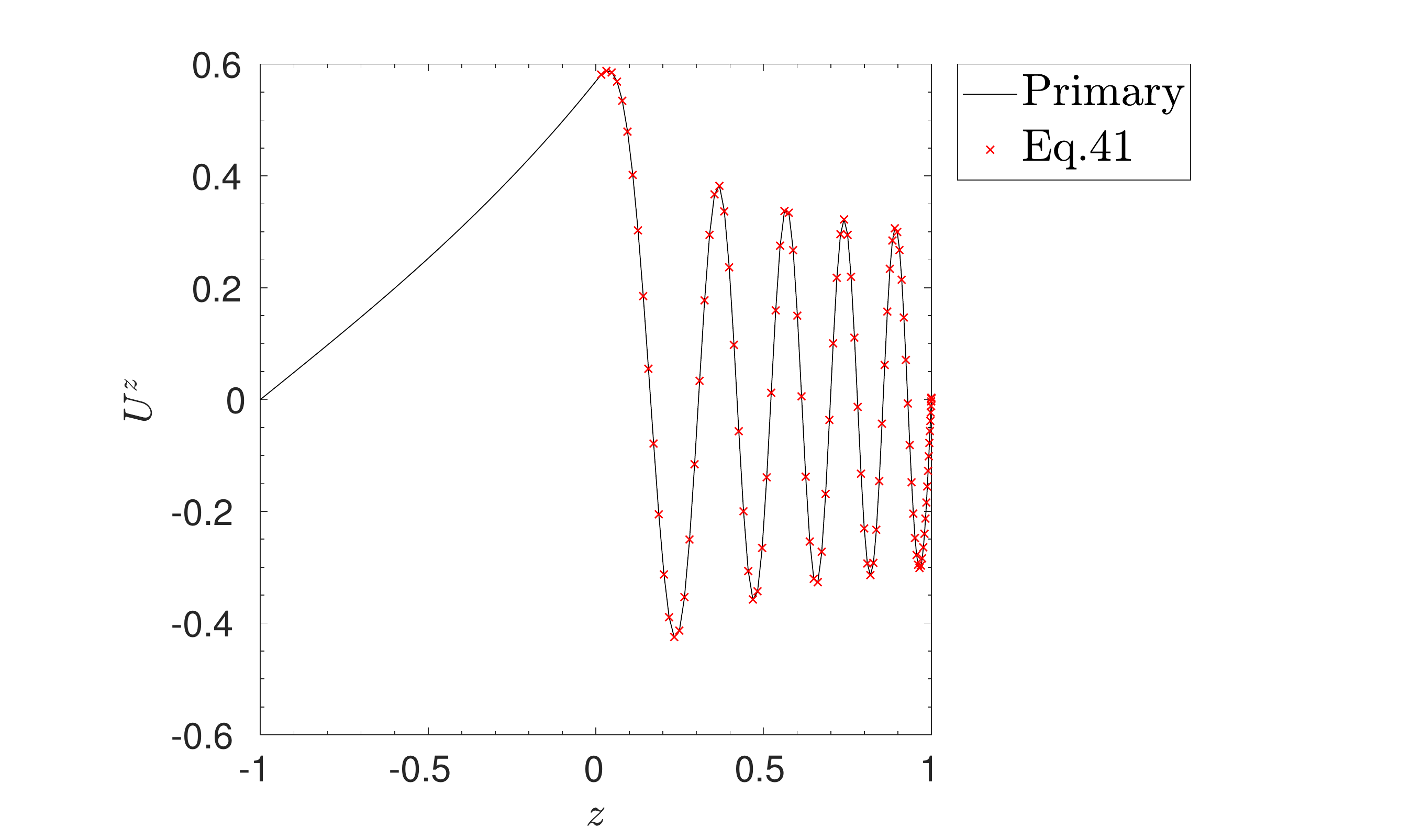}\label{4a}}
      \subfigure[Secondary $U^z$ vs $z$ at $t=5000$.]{\includegraphics[trim=4cm 0cm 4cm 0cm, clip=true,width=0.49\textwidth]{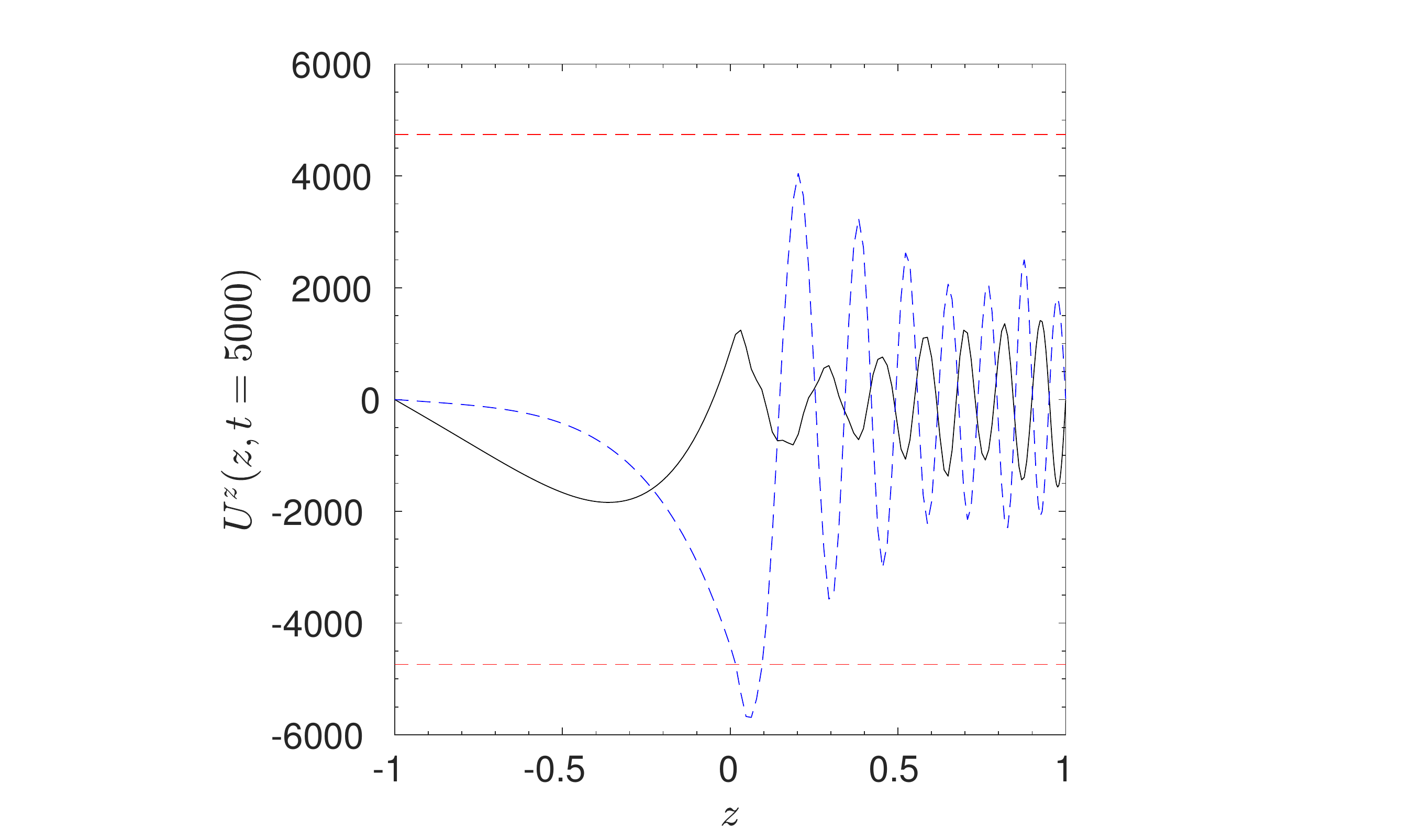}\label{4b}}
      \subfigure[Secondary $U^z$ vs $t$ at $z=0.1$.]{\includegraphics[trim=4cm 0cm 5cm 0cm, clip=true,width=0.49\textwidth]{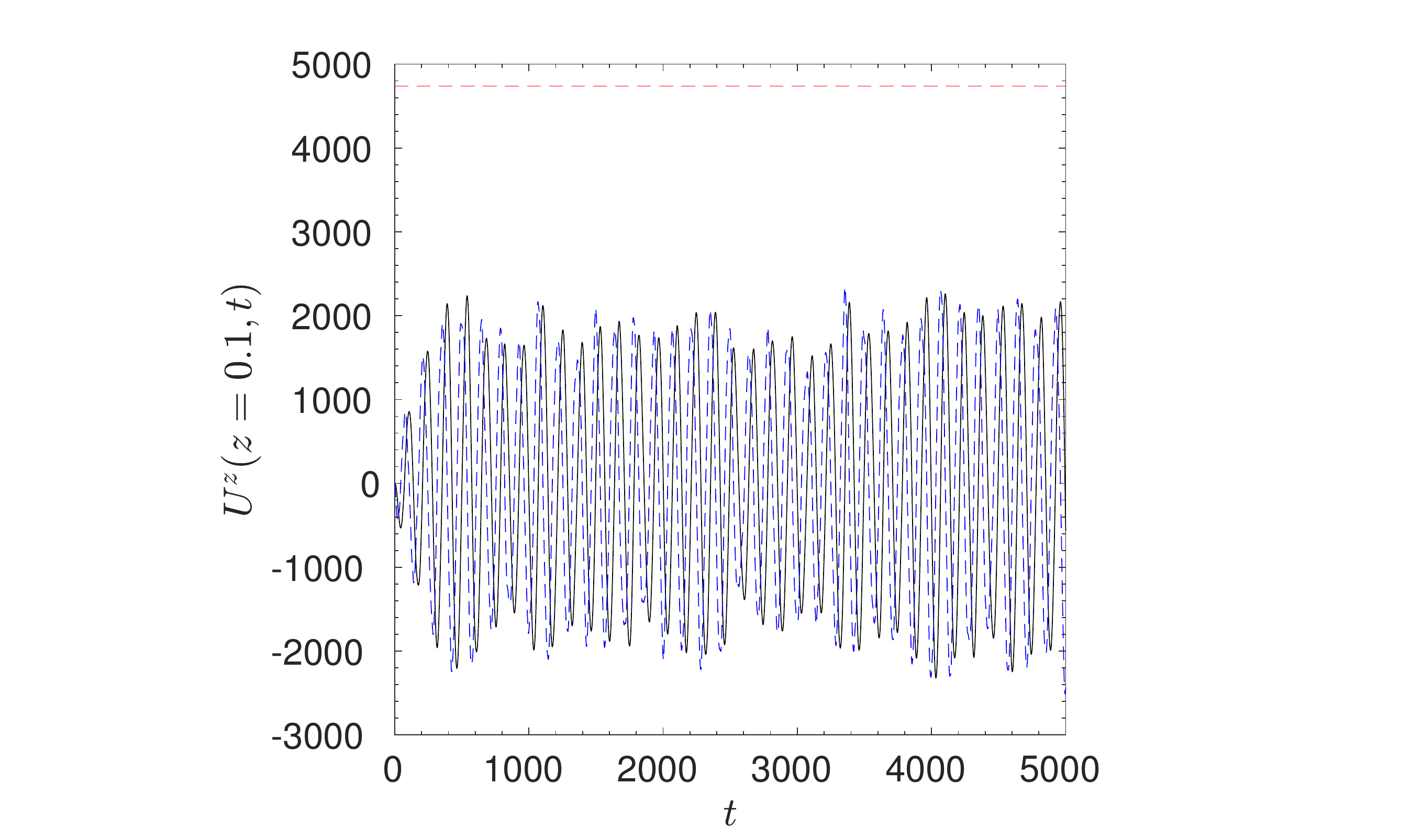}\label{4c}}
      \subfigure[Frequency power spectrum of $U^z(z=0.1,t)$.]{\includegraphics[trim=3cm 0cm 4cm 0cm, clip=true,width=0.49\textwidth]{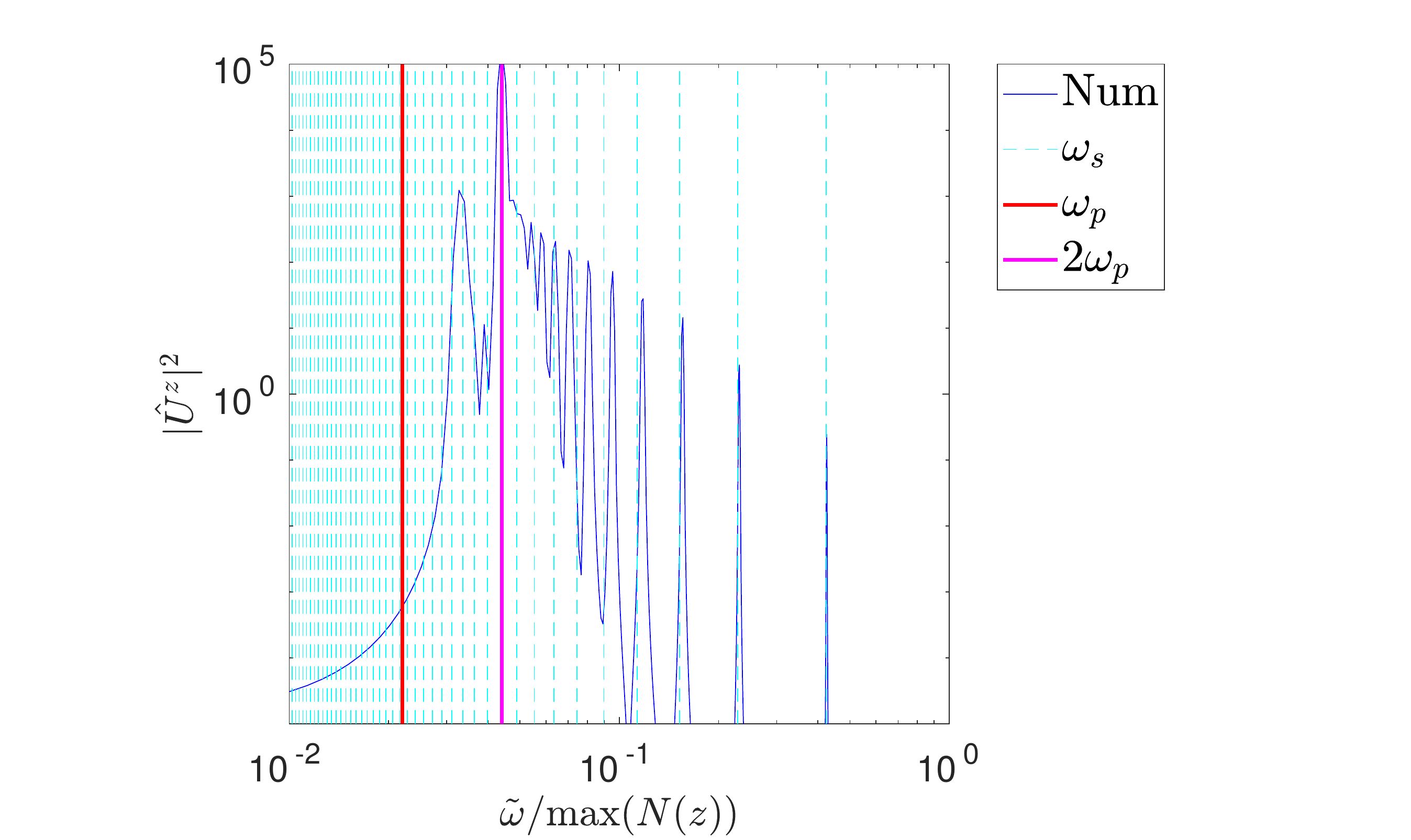}\label{4d}}
    \end{center}
  \caption{Same as Fig.~\ref{Fig2} but for a mode with $\omega_p=\mathrm{Re}[\omega_i]\approx0.0220$.}
  \label{Fig4}
\end{figure*}

We first show the spectrum of $\omega_i$ from solving the above eigenvalue problem in the complex plane in Fig.~\ref{eval}, where we highlight a particular low frequency mode with $\omega_p\approx0.0456$ with a red star. We show the analytical predictions for $\omega_i$ from (\ref{mp5}) as the vertical cyan dashed lines, and our numerical values for the complex mode frequencies for the modes with $k=1$ as black crosses and $k=2$ as blue circles (the latter shows the ``secondary" modes that can be excited). The agreement is excellent between (\ref{mp5}) and our numerical results for the $k=1$ modes for $\mathrm{Re}[\omega_i]\lesssim0.1$, with a slight departure, as expected, when the frequency is larger. 

In Fig.~\ref{2a}, we show the $U^z$ eigenfunction for the particular primary mode with $\omega_p\approx0.0456$ together with the analytical prediction in (\ref{b20}) in the stable layer, again indicating excellent agreement. We then show in Figs.~\ref{2b} --\ref{2d} the spatial structure of the secondary solution at $t=5000$ (the real and imaginary parts shown as solid and dashed lines, respectively), the temporal evolution of the secondary $U^z$ at $z=0.1$, and the frequency power spectrum ($|\hat{U}^z|^2$) of the secondary signal $U^z(z=0.1,t)$. By $t=5000$, modes with multiple frequencies still contribute to the signal, though the power is maximised in the mode with a frequency close to $2\omega_p$. In Fig.~\ref{2d}, we also show the predicted frequencies of the secondary modes $\omega_s$ from solving the above eigenvalue problem in the case $k=2$, which indicates that many of these modes are excited and have not yet fully damped, as we expect from our initial conditions. Figs.~\ref{2b} --\ref{2d} show that the analytical prediction for the secondary mode amplitude $a_s$ in (\ref{mp14}), which is shown as the red horizontal lines in Figs.~\ref{2b} and \ref{2c}, agrees quite well with our numerical results in this case. The secondary mode amplitudes are dominant near $z\sim 
0$, where the primary mode also takes its maximum value. We observe here that, due to the near-resonance explained in \S~\ref{weaklynonlin}, the amplitude $U^z$ of the super-harmonic secondary solution can far exceed that of the primary mode, and hence we predict this nonlinear effect to be important in this system because $a_s\gg C_p$.

We also show results for a larger frequency mode with $\omega_p\approx0.263$ in Figs.~\ref{3a}--\ref{3d} and a lower frequency mode with $\omega_p\approx0.0220$ in Figs.~\ref{4a}--\ref{4d}, with otherwise the same parameters. The primary wave solution (and its frequency) for the larger frequency case is clearly described less well by our analytical prediction, but the agreement is still good. The amplitude predicted for $a_s$ is also comparable with the values we observe numerically in both of these examples. 

Our numerical calculations here have thus verified the analytical results derived for low frequencies in \S~\ref{boussi}--\ref{weaklynonlin}, so we can be more confident in their application. It would be interesting to explore the fully nonlinear evolution of the system studied here using numerical simulations to analyse the long-term dynamics of the generation of super-harmonic secondary waves and their interaction with the primary (tidal) wave. These simulations will be presented in a future publication.

\section{Application of the criteria for transition to nonlinearity to main-sequence stellar models}
\label{application}

\begin{figure}
\includegraphics[width=0.5\textwidth]{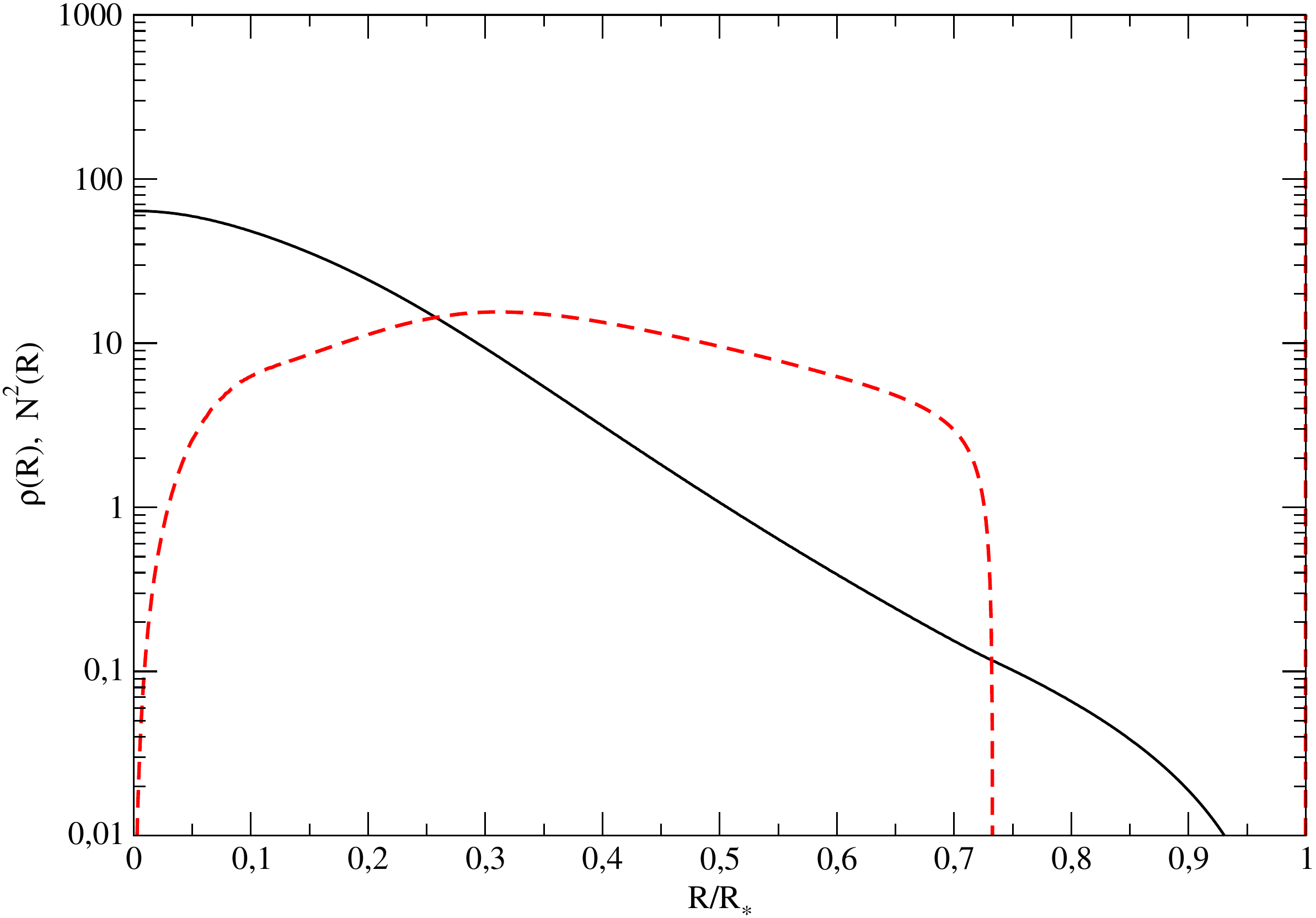}
\caption{Radial profiles of stellar density $\rho$ (black solid line) and square of the Brunt-V\"{a}is\"{a}l\"{a} frequency $N^2$ (red dashed line), both in the natural units defined in \S~\ref{dense}, for model 1a.}
\label{M1}
\end{figure}

\begin{figure}
\includegraphics[width=0.5\textwidth]{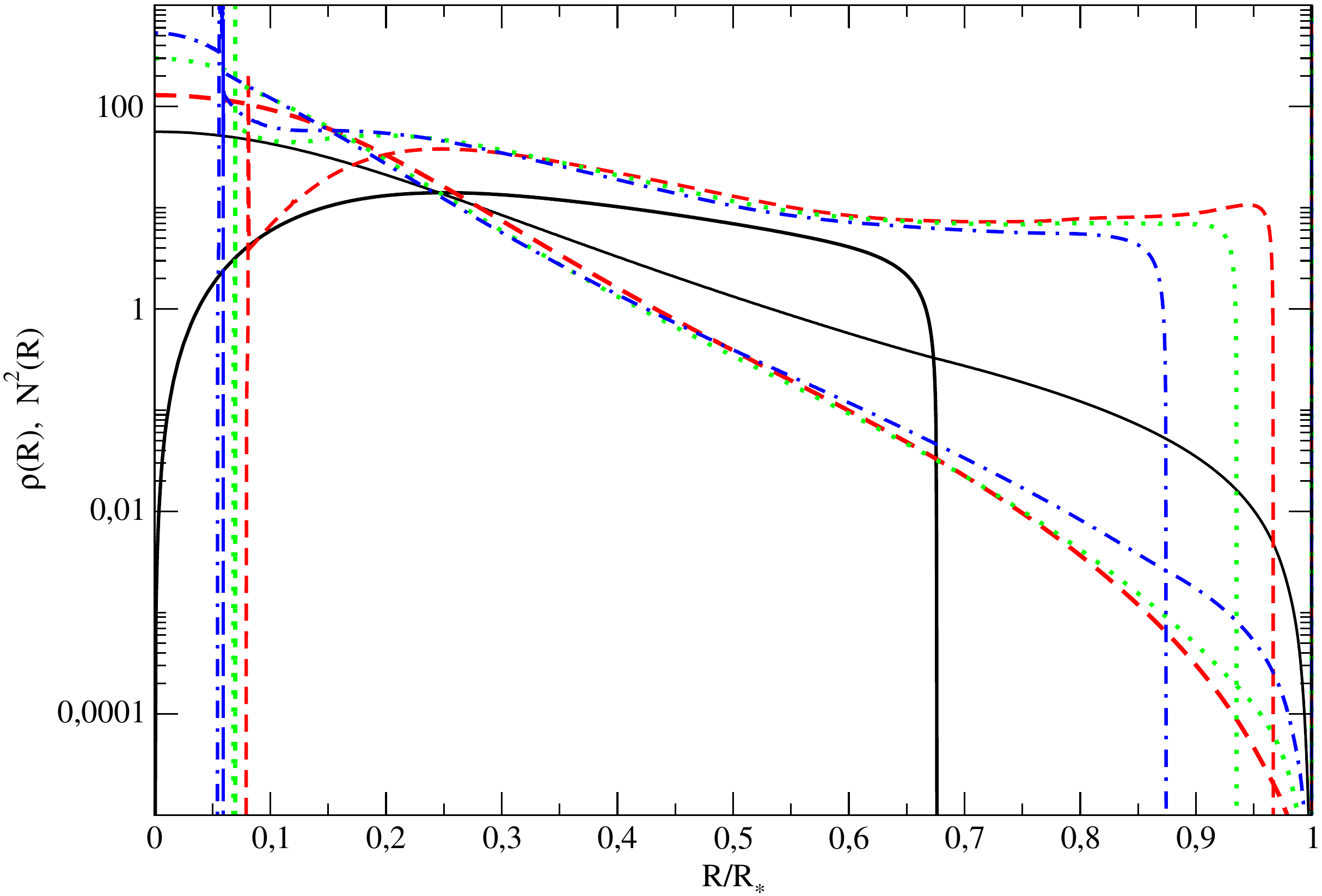}
\caption{Radial profiles of $\rho $ and $N^2$ in natural units for the models 1.4a, 1.4b, 1.4c and 1.4d shown as black solid, red dashed, green dotted and red dot-dashed lines, respectively. Although the lines corresponding to  $\rho $ and $N^2$ calculated for the same model have the same style, they can be easily distinguished from each other, since $\rho$ decreases monotonically with $R$.}
\label{M1.4}
\end{figure}

\begin{figure}
\includegraphics[width=0.5\textwidth]{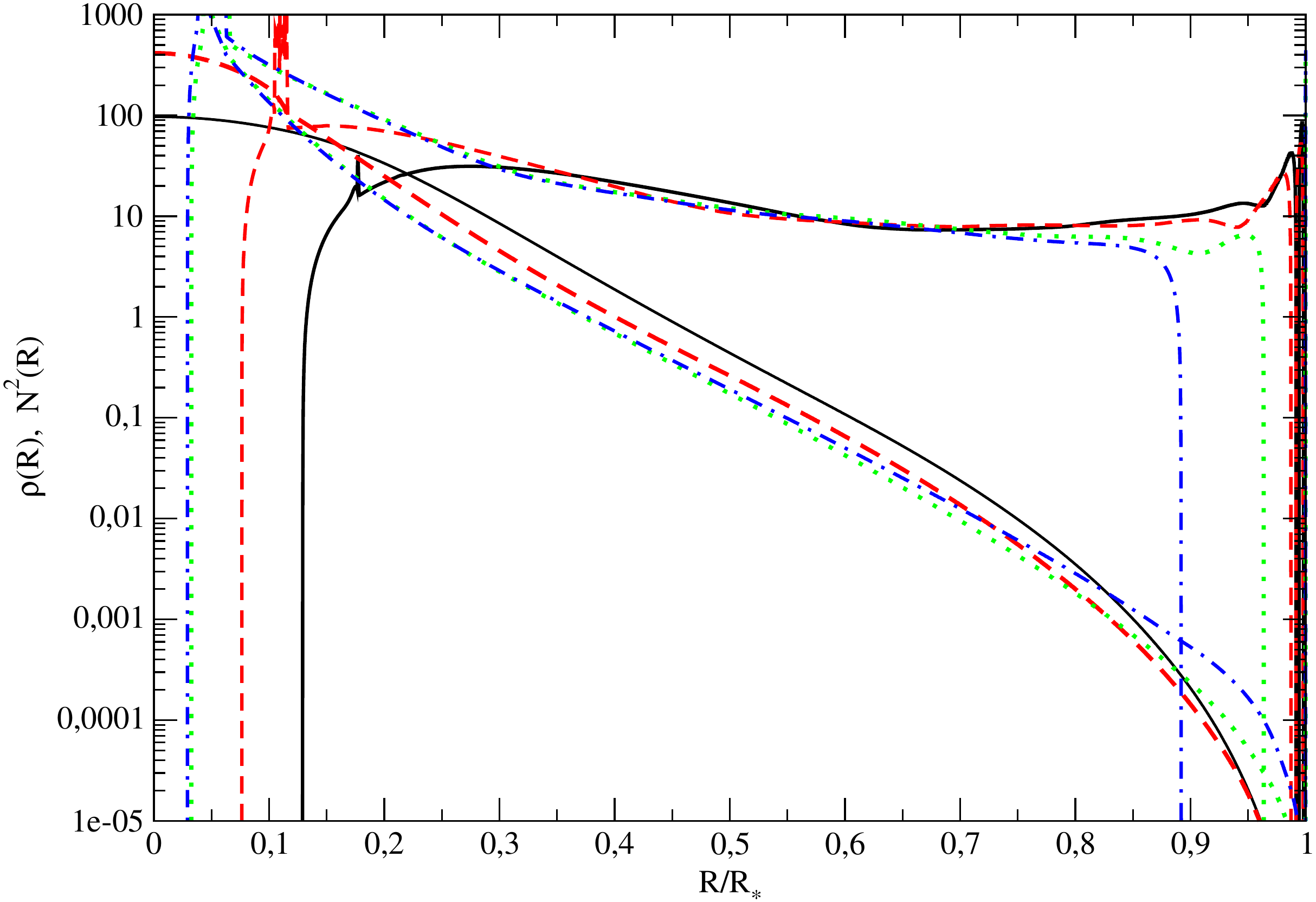}
\caption{Same as Fig. \ref{M1.4}, but for the models 2a, 2b, 2c and 2d shown as black solid, red dashed, green dotted and blue dot-dashed lines, respectively.}
\label{M2}
\end{figure}

In this section, we turn to apply our criteria for the transition to nonlinearity in several main-sequence stellar models computed using MESA version 15140\footnote{See \url{http://mesa.sourceforge.net}.} \citep{Paxton2011,Paxton2013,Paxton2015,Paxton2018,Paxton2019}. We calculate the values of $C^{crit}$ for each of the various scenarios given by eqns. (\ref{e12}), (\ref{core2}), (\ref{mp22}) and (\ref{mp26}) for a number of main-sequence stellar models with masses $M=1$, $1.4$ and $2M_{\odot}$ and different ages. The basic characteristics of our models are given in table \ref{param}, where we show stellar masses in units of $M_{\odot}$, their radii in units of $R_{\odot}$, ages in years, positions of the base of the convective zone, $r_c$, in units of the stellar radius $R_*$, density at the base of convective zone, $\rho_c$, in units of the mean density $\bar \rho = 3M_\ast/(4\pi R_\ast^3)$, and the mean density in units of the solar mean density, and finally the values of $\omega_c$ and $I$ in units of $\sqrt{GM_{\ast}\over R_{\ast}^3}$, respectively. Additionally, we show radial profiles of the Brunt-V\"{a}is\"{a}l\"{a} frequency and stellar density in Figs. \ref{M1}, \ref{M1.4} and \ref{M2}, respectively.    

\begin{table*}
\caption{Basic characteristics of our stellar models}
 \begin{tabular}{ccccccccc}
 \hline
  Model & Mass ($M_\odot$) & Radius ($R_\odot$) & Age (years) & $r_c$ ($R_\ast$) & $\rho_c/\bar \rho$ & $\bar \rho$  &$\omega_c$ & $I$\\
  \hline
  1a & 1   & 0.96 & $2\times10^9$    & 0.73 & $0.12$ & 1.13 & 9.36 & 9.42\\
  1.4a & 1.4 & 1.65 & $1.07\times10^7$ & 0.68 & $0.32$ & 0.31 & 9.01 & 8.52\\
  1.4b & 1.4 & 1.43 & $2.03\times10^8$ & 0.97 & $2.05\times10^{-5}$ & 0.48 & 49.2 & 10.3\\
  1.4c & 1.4 & 1.72 & $1.73\times10^9$ & 0.94 & $1.89\times10^{-4}$ & 0.28 & 30.7 & 14.3\\
  1.4d & 1.4 & 1.93 & $2.38\times10^9$ & 0.87 & $2.59\times10^{-3}$ & 0.19 & 20.5 & 19.0\\
  2a & 2 & 1.63 & $2.67\times10^7$ & 0.99 & $1.78\times10^{-7}$ & 0.46 & 182.3 & 8.07\\
  2b & 2   & 2.68 & $9.97\times10^8$ & 0.99 & $5.94\times10^{-7}$ & 0.10 & 113.4 & 16.7\\
  2c & 2 & 5.35 & $1.41\times10^9$ & 0.96 & $3.02\times10^{-5}$ & 0.013 &  41.6 & 52.5\\
  2d & 2 & 5.75 & $1.44\times10^9$ & 0.89 & $6.09\times10^{-4}$ & 0.011 &  17.7 & 59.9\\
  \hline
 \end{tabular}
\label{param}
\end{table*}


\begin{figure}
\includegraphics[width=0.49\textwidth]{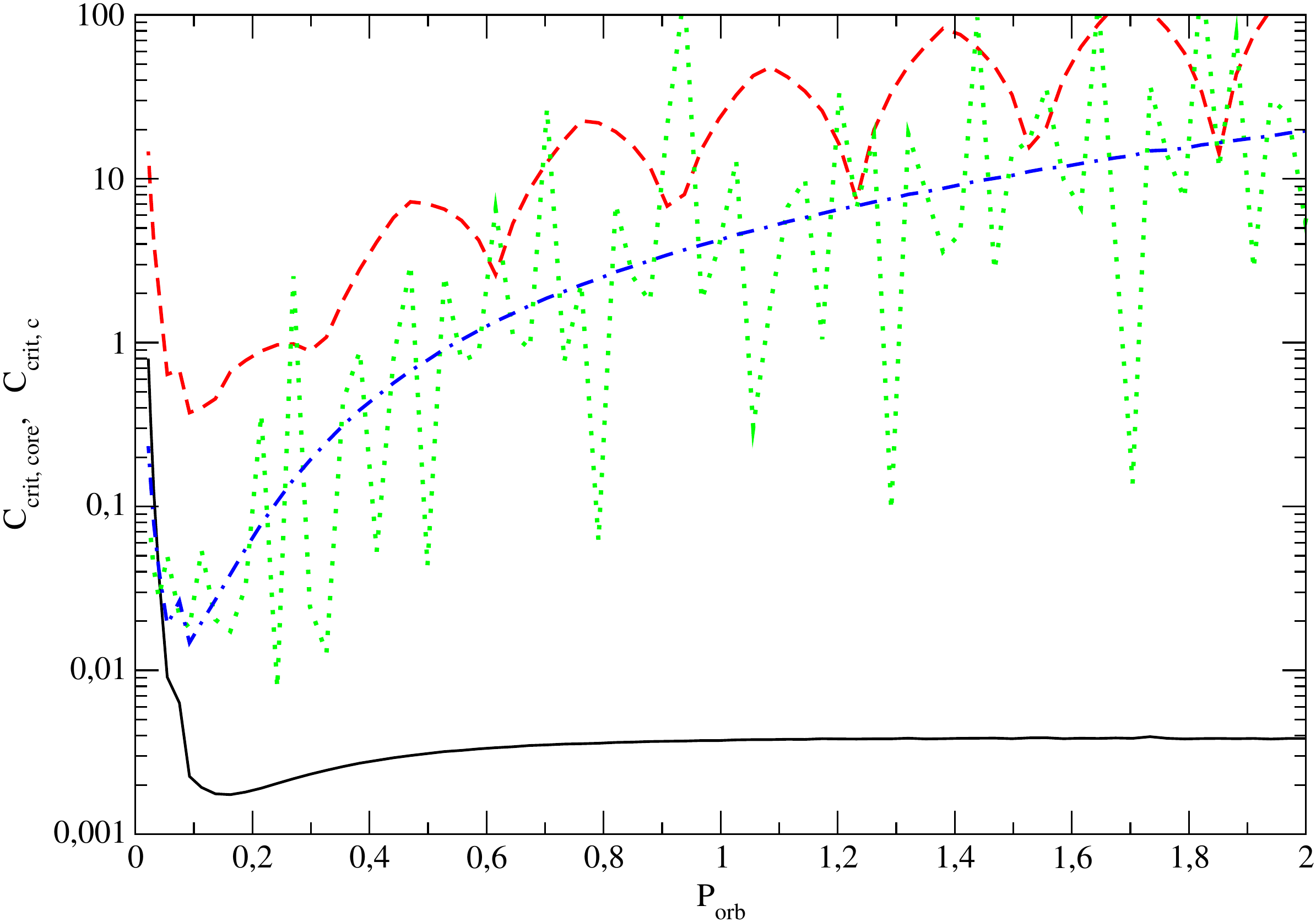}
\caption{Various criteria for transition to nonlinearity $C_{crit}$ are shown for a solar mass model as a function of orbital period in days. The black solid curve corresponds to $C_{centre}$ using the `standard' expression (\ref{e12}) for wave breaking in radiative cores, the red dashed one corresponds to $C_{crit,c}$ in (\ref{mp22}), the green dotted and blue dot-dashed curves are given by $C_{crit,c}^{dense}$ in eq. (\ref{mp26}), where in the former we set $\kappa=0$ and in the latter $\kappa=1$.}
\label{cr1}
\end{figure}

In Fig.~(\ref{cr1}) we show the quantities $C_{centre}$, $C_{crit, c}$, and $C_{crit, c}^{dense}$ for both $\kappa=0$ 
and $\kappa=1$, as given by equations (\ref{e12}), (\ref{mp22}) and(\ref{mp26}), respectively, as functions of orbital period for the solar model. We observe that according to these criteria, the transition to non-linearity at the stellar centre always happens for much smaller mass ratios than any of the other measures i.e. $C_{centre} \ll C_{crit,c}, C_{crit,c}^{dense}$. This criterion is equivalent to the condition for nonlinearity discussed by \cite[e.g.][]{GD98,OL2007,BO2010,Barker2020}, and is the most important measure of nonlinearity in our solar-mass model. 

Note that in order for our criteria to predict the onset of nonlinearity, due to its definition we require $C_{crit} \lesssim 1$ for any mass ratio.
From Fig. (\ref{e12}), we see that $C_{crit, c}$ given by (\ref{mp22}) becomes smaller than one only for very short orbital periods $P_{orb} \sim 0.2\mathrm{d}$ and the corresponding mass ratio should be larger than $\sim 0.6$ i.e.~nonlinearity is predicted only for comparable mass ultra-close binaries. Assuming that when our criteria for the transition to non-linearity are satisfied the primary mode is then efficiently damped, we conclude here that generation of super-harmonic secondary modes near the radiative-convective interface in our solar model is unlikely to provide an important contribution to tidal dissipation.

It is interesting to note that both $C_{crit,c}$ and $C_{crit,c}^{dense}(\kappa = 0)$ are oscillatory functions of orbital period. In the former case, this is due to the oscillatory character of the function $f(n)$ shown in Fig.~\ref{fn}. In the latter case, the behaviour is even more strongly oscillatory, which is due both to the oscillation of $f(n)$ with $n$ and to the sharp periodic changes of $S(\delta, \kappa)$ with $\delta$ when $\kappa=0$, as indicated by eq. (\ref{l7}). These curves have the same oscillatory behaviour in the more massive stellar models considered below.

\begin{figure*}
  \begin{center}
      \subfigure{\includegraphics[trim=0cm 0cm 0cm 0cm, clip=true,width=0.49\textwidth]{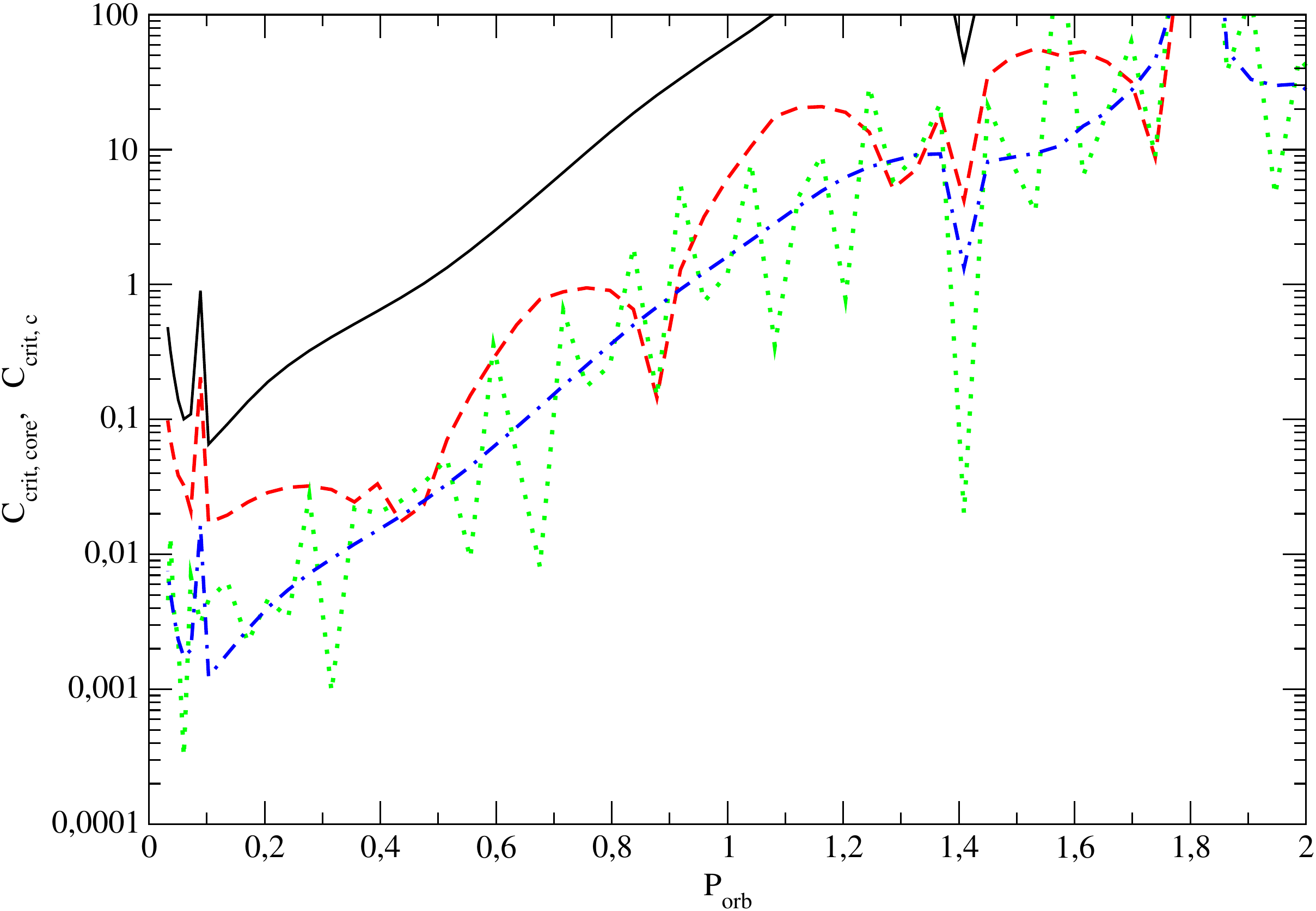}}
      \subfigure{\includegraphics[trim=0cm 0cm 0cm 0cm, clip=true,width=0.49\textwidth]{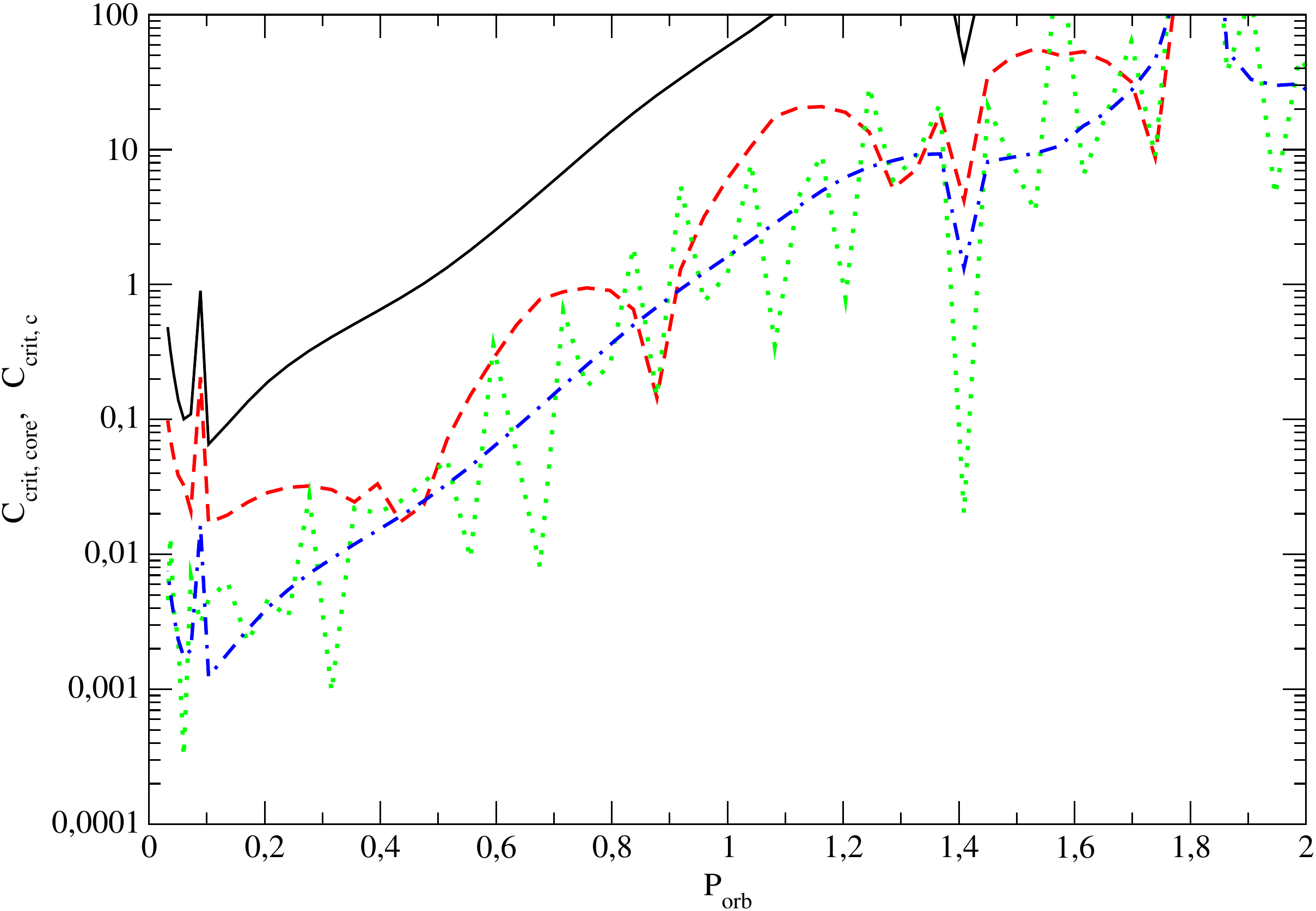}}
      \subfigure{\includegraphics[trim=0cm 0cm 0cm 0cm, clip=true,width=0.49\textwidth]{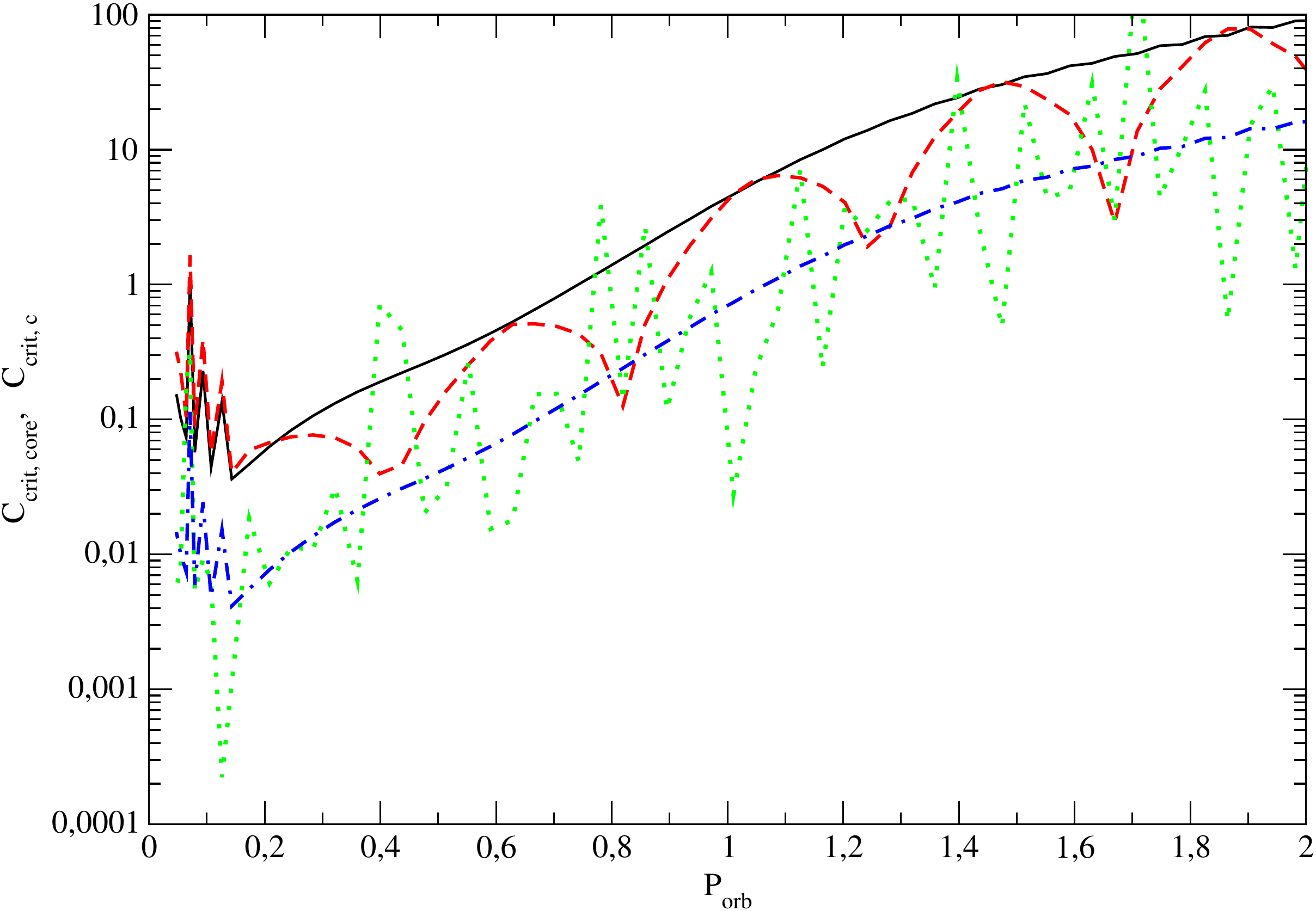}}
      \subfigure{\includegraphics[trim=0cm 0cm 0cm 0cm, clip=true,width=0.49\textwidth]{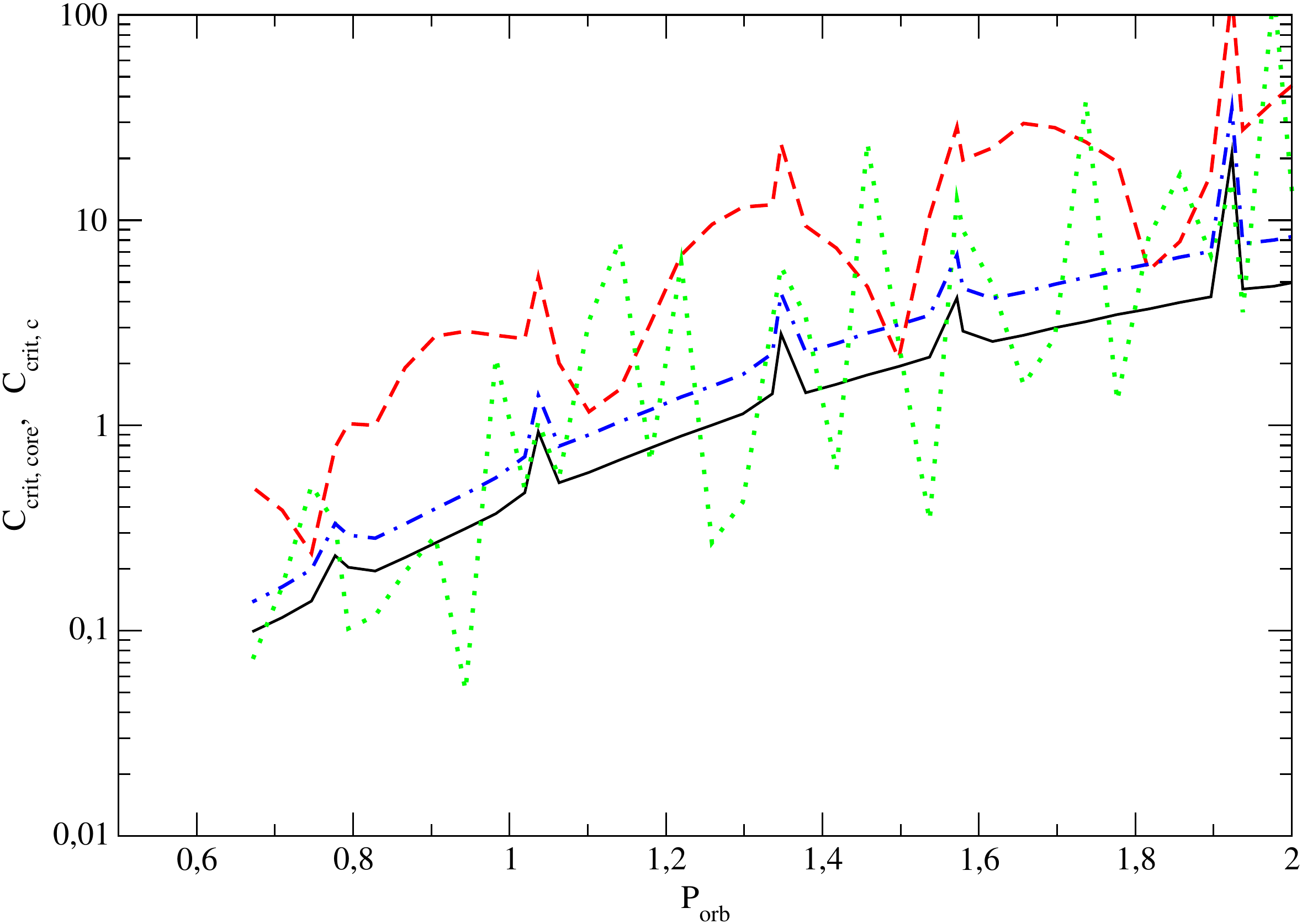}}
    \end{center}
  \caption{Same as Fig.~\ref{cr1}, but for our models with $M=1.4M_{\odot}$, and the black solid line shows (\ref{core2}). Top left, top right, bottom left and bottom right panels correspond to the models a, b, c, d, respectively.}
  \label{cr2}
\end{figure*}

\begin{figure*}
  \begin{center}
      \subfigure{\includegraphics[trim=0cm 0cm 0cm 0cm, clip=true,width=0.49\textwidth]{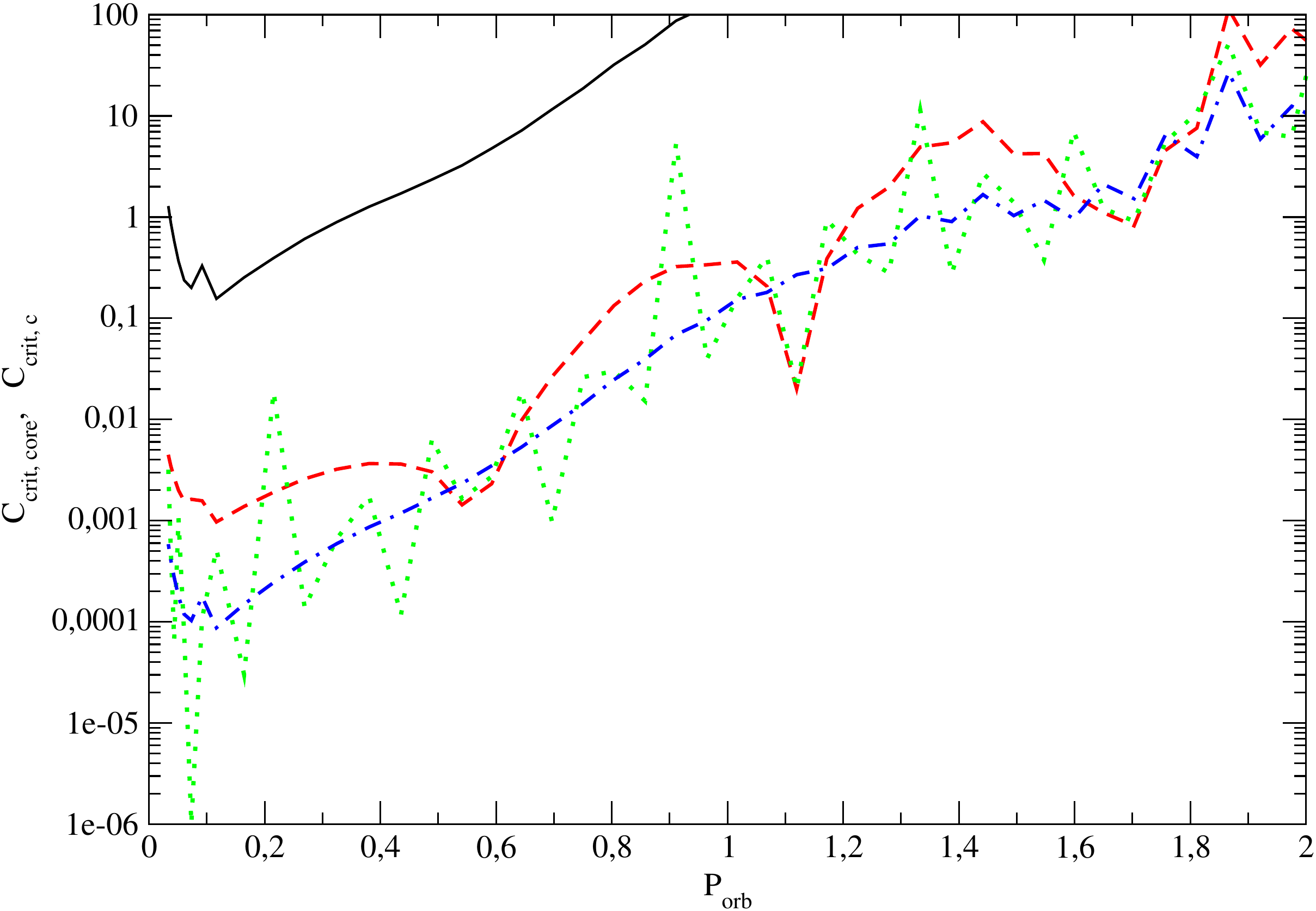}}
      \subfigure{\includegraphics[trim=0cm 0cm 0cm 0cm, clip=true,width=0.49\textwidth]{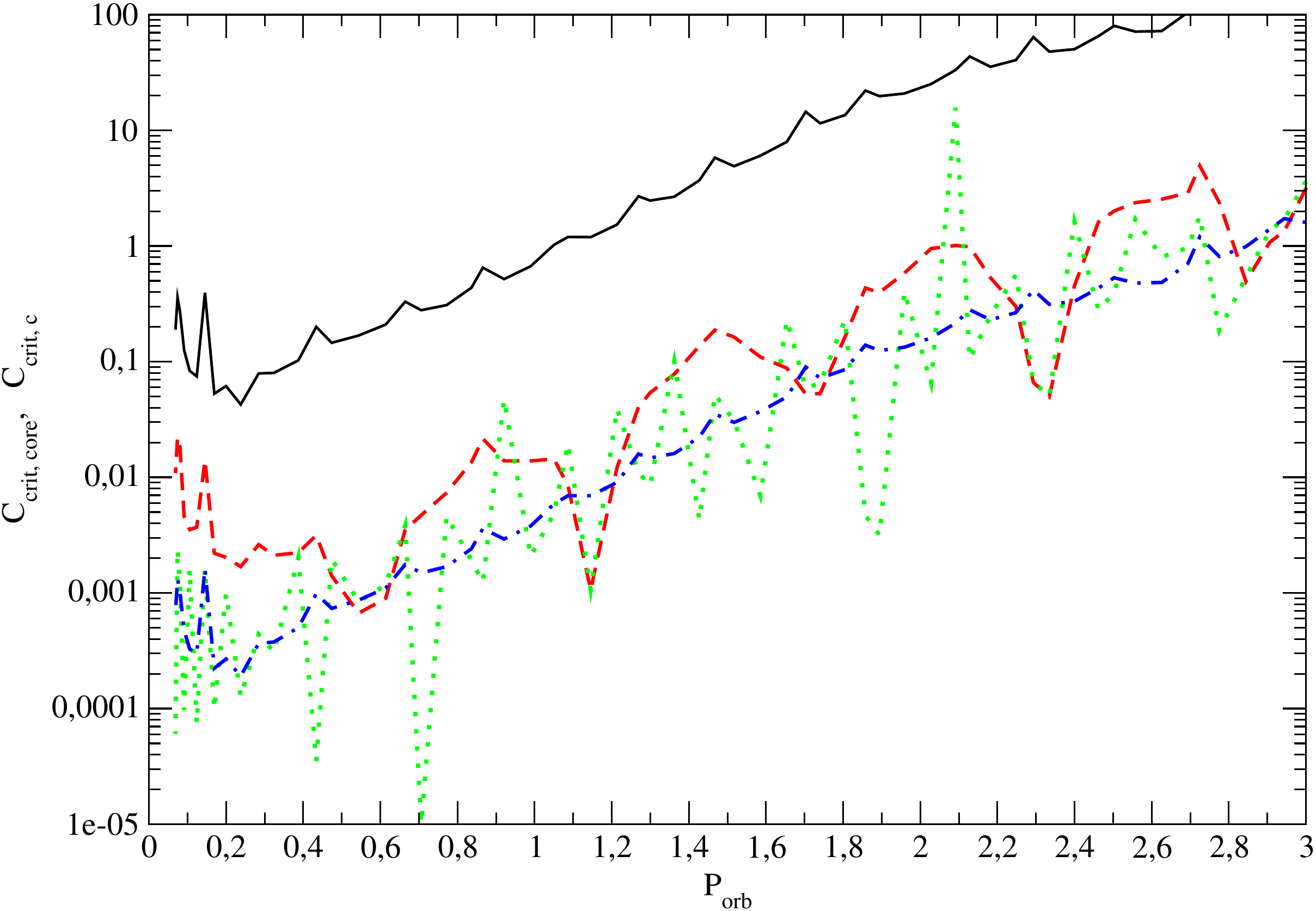}}
      \subfigure{\includegraphics[trim=0cm 0cm 0cm 0cm, clip=true,width=0.49\textwidth]{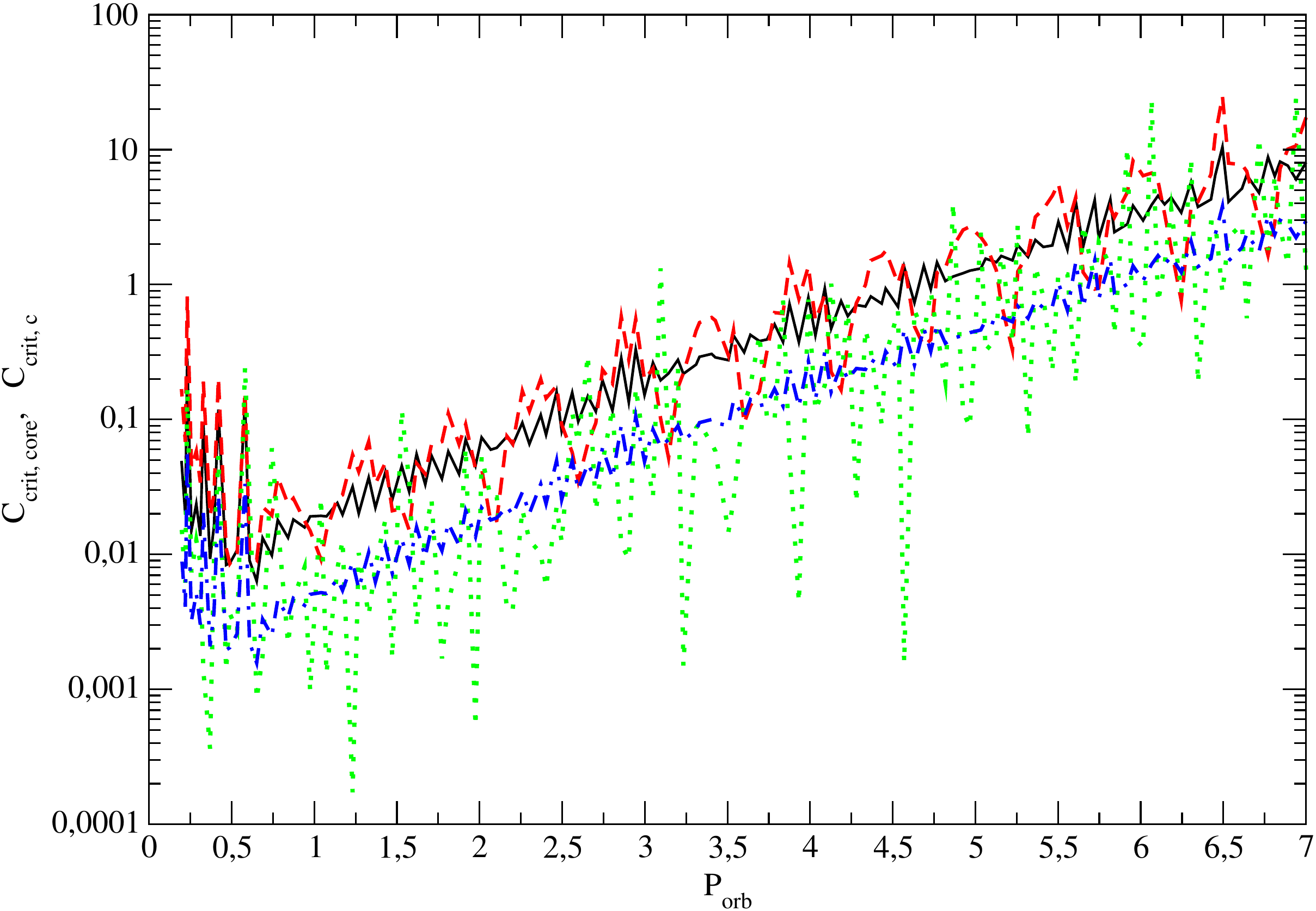}}
      \subfigure{\includegraphics[trim=0cm 0cm 0cm 0cm, clip=true,width=0.49\textwidth]{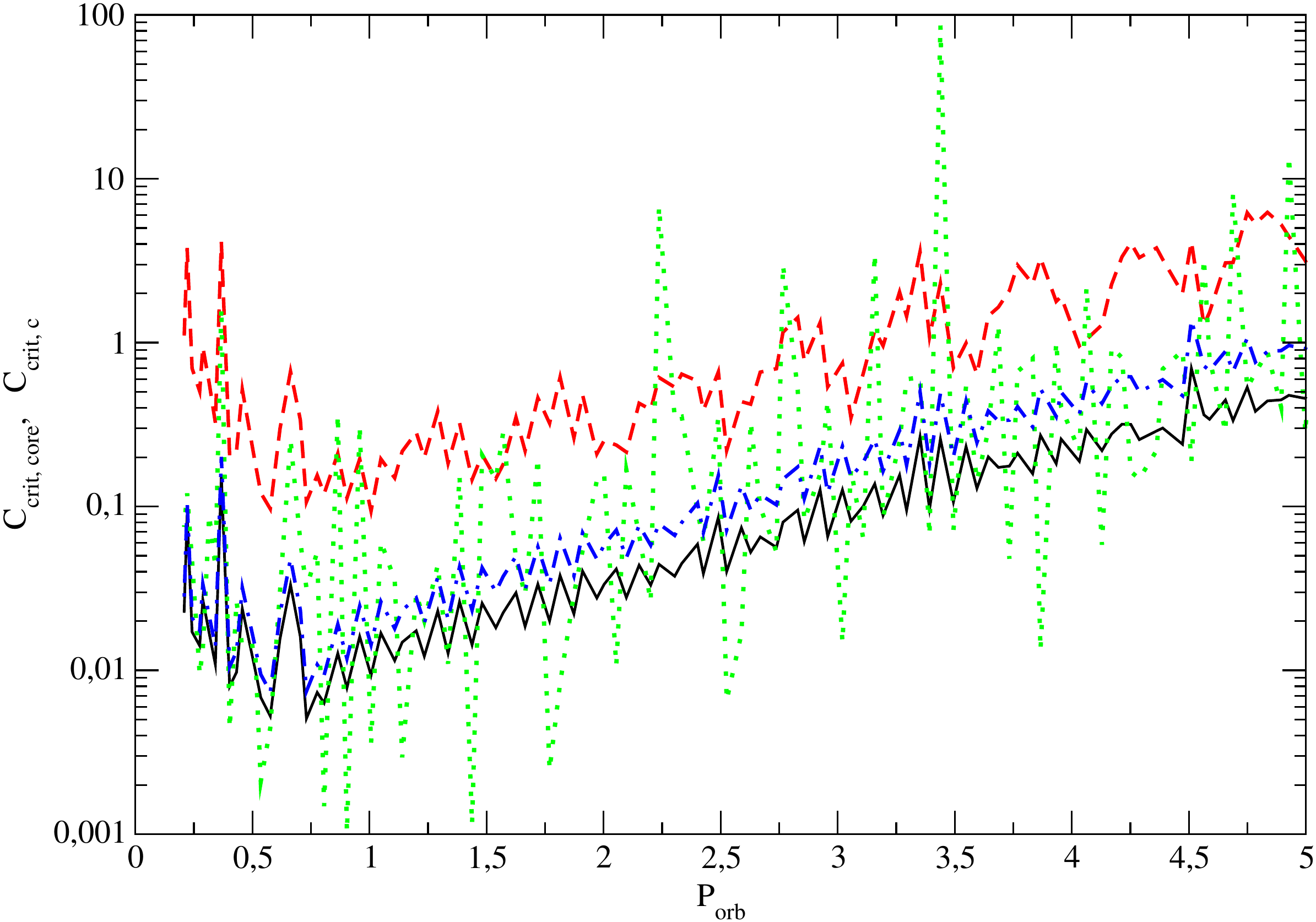}}
    \end{center}
  \caption{Same as Fig. \ref{cr2}, but for the models with $M=2M_{\odot}$.}
  \label{cr3}
\end{figure*}

\begin{figure}
\includegraphics[width=0.49\textwidth]{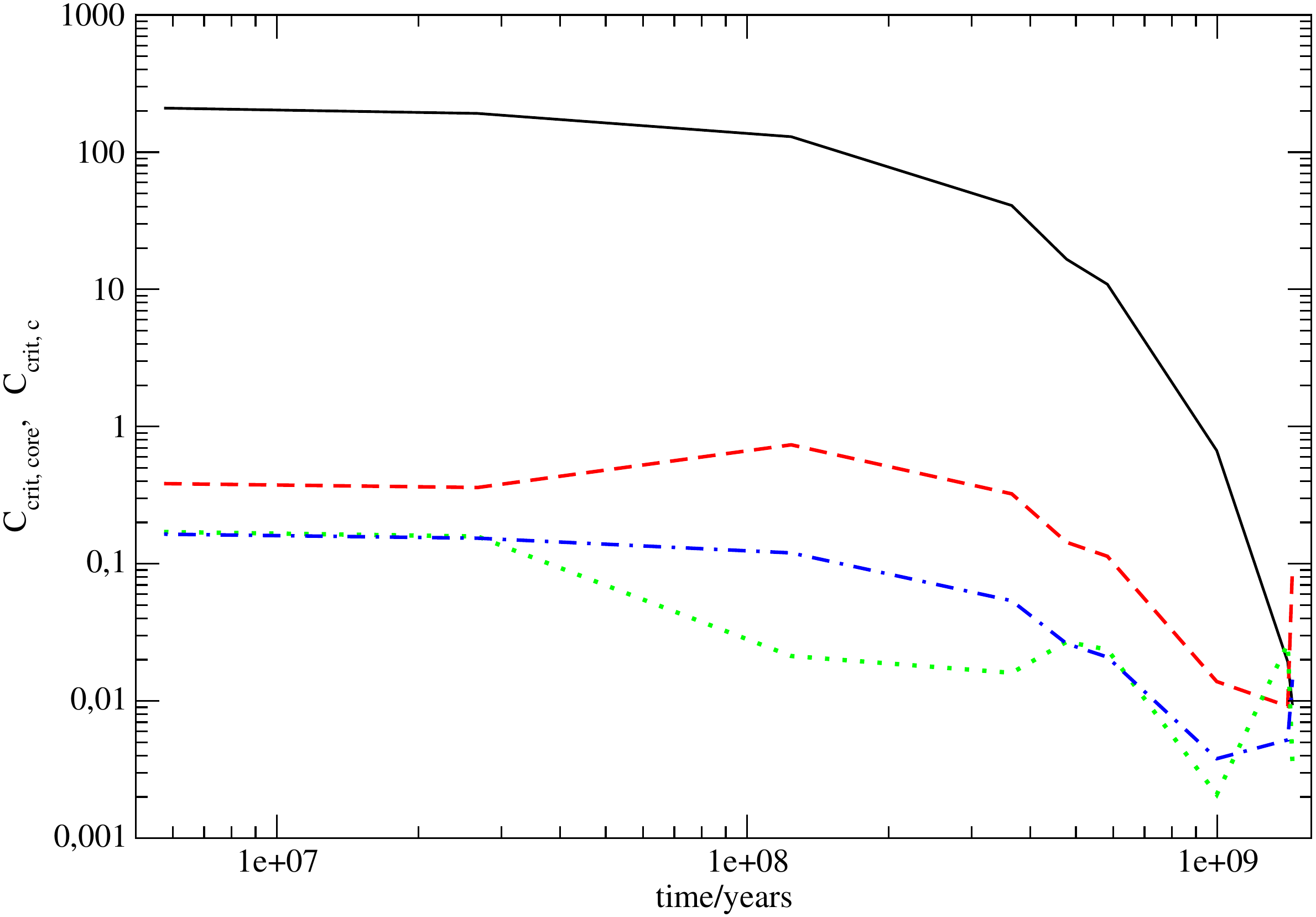}
\caption{Various expressions for $C_{crit}$ as a function of stellar age in years for stellar models with $M=2M_{\odot}$. Solid curve,
dashed, dotted and dot-dashed curves correspond to (\ref{core2}), (\ref{mp22}), (\ref{mp26}) with $\kappa=0$ and (\ref{mp26}) with $\kappa=1$, respectively.}
\label{Ct}
\end{figure}

Figs.~(\ref{cr2}) and (\ref{cr3}) are similar to Fig.~(\ref{cr1}), but the curves are calculated for more massive
models with $M=1.4M_{\odot}$ and $M=2M_{\odot}$. Contrary to our solar model, apart from model 
1.4a the more massive stars have convective cores, as is shown in Figs.~\ref{M1.4} and \ref{M2}. In this case, the criterion for wave breaking near convective cores predicted by (\ref{core2}) must be used instead of the criterion for wave breaking at the centre of a radiative core in (\ref{e12}). The corresponding curve is shown as a solid line in all figures apart from the one corresponding to 1.4a, where (\ref{e12}) is shown in the same way instead. Unlike the solar case we see from (\ref{cr2}) and (\ref{cr3}) that 
the criteria for transition to nonlinear behaviour at the outer convective/radiative interface is satisfied for smaller mass ratios than 
the criterion for transition at the convective core (\ref{core2}), for almost all stellar ages and orbital periods\footnote{We ignore the very large amplitude oscillations of $C^{dense}_{crit, c}$ due to the trigonometric nature of $S(\delta, \kappa=0)$ and assume that only some average value of this quantity has physical meaning.}. The exceptions are models 1.4a, 1.4d and 2d. In the case of 1.4a, we 
see from Fig.~\ref{M1.4} that this model has a radiative core and an extended convective envelope. This model is rather similar to the solar model and it is not surprising that it demonstrates similar behaviour. The models 1.4d and 2d also have quite large convective envelopes. The larger the transition radius to the convective envelope, $r_c$, the smaller is the density at this radius, $\rho_c$, and the larger is typical frequency, $\omega_c$, at this radius. Both the smaller density and larger $\omega_c$ values lead to the smaller values of $C_{crit, c}$ seen in most of these models. Note, however, that there are two other important factors determining values of $C_{crit, c}$, namely, values of the overlap integrals in units of the frequency $\Omega_*=\sqrt{GM_*\over R_*^3}$, and a value for the ``average frequency" of a star. The former factor leads to smaller $C_{crit, c}$ for stars with extended convective envelopes, since the overlap integrals are typically larger for such stars \citep[e.g.][]{ChPI2013,ChPI2017}. Smaller values of the average density lead to larger values of the tidal forcing frequency expressed in units of the natural frequency and, as a result, smaller values of $C_{crit, c}$. It may be possible that these two factors counterbalance each other for certain stellar models however.

If we speculate that whenever our criteria for transition to nonlinear behaviour are satisfied, the primary tidally-excited modes are efficiently damped, our results concerning more massive models could have important implications for tidal evolution around such stars. They would imply that above a critical mass ratio, tidal dissipation rates can be straightforwardly computed as a known function of orbital period, stellar mass and age (essentially according to linear tidal theory assuming a fully damped/MLD/travelling wave regime). 
In some cases, this ratio can be as small as $\sim 10^{-3}$ for $P_{orb}\sim 1\mathrm{d}$, as in e.g. models 2a-2c, indicating that planetary mass companions can cause a strongly nonlinear tidal response in the star, potentially with efficient tidal dissipation. To illustrate this further, we show the dependence of $C_{crit, core}$, $C_{crit, c}$ and $C_{crit, dense}$ on stellar age in Fig.~\ref{Ct}, keeping the orbital period fixed at one day.  As shown in this figure, typical values of  $C_{crit, dense}$ are indeed as small as $10^{-3}$ for stellar ages $\sim 10^{9}\mathrm{yr}$, indicating that this mechanism can potentially be important for tidal evolution of hot Jupiters around such stars.

\section{Conclusions and Discussion}
\label{conclusions}

We have studied the role of nonlinear effects on tidally-excited gravity waves in the radiation zones of stars, primarily focussing on a new mechanism that could be important in stars possessing convective cores. Our work was partly motivated to study tides due to massive short-period hot Jupiters, which are observed to preferentially orbit stars with convective cores \citep[e.g. WASP-18 b][]{Wilkins17}. For these stars, the geometric focussing and consequent breaking (or nonlinear wave-wave interactions) of gravity waves in the stellar core (which we have revisited in \S~\ref{WaveBreakCore}), which can result in efficient tidal dissipation when this occurs \citep[e.g.][]{GD98,OL2007,BO2010,Barker2011,BO2011,W12,EW16,Barker2020}, cannot take place for planetary mass companions, unlike in solar-type with radiative cores. 

We have developed a theory for the nonlinear excitation of super-harmonic `secondary' internal gravity waves by a `primary' gravity wave, assuming that the latter is generated by tidal forcing in a main-sequence star due to an orbiting companion. This excitation appears to be the most efficient near the transition radius, $r_c$, between a radiative interior and a convective envelope. At this location, the usual WKBJ approximation for the description of high radial order gravity waves is strictly invalid. Similar to the analogous problem recently studied in fluid dynamics and oceanography \citep[see e.g.][and references therein]{Wunsch2017,Baker2020}, the nonlinear self-interaction of an internal gravity wave propagating in a region with a spatially-varying Brunt-V\"{a}is\"{a}l\"{a} frequency can generate super-harmonic secondary waves in stars, which can be in near-resonance with the primary mode and potentially be excited to large amplitudes. If this mechanism operates, this implies that nonlinear effects could be important for tidal waves even in stars with convective cores with planetary mass companions (as we have estimated in \S~\ref{application}).

We have adopted a number of simplifying assumptions to model the generation of super-harmonic secondary waves. Firstly, we have considered a region of small spatial extent near $r=r_c$, and employed a local Cartesian model instead of global spherical geometry. Secondly, we assumed the square of the Brunt-V\"{a}is\"{a}l\"{a} frequency to be a linear function of the distance from $r_c$ into the radiative layer \citep[see e.g.][for a justification and discussion of the limitations of this assumption]{Barker2011,IPCh}. Thirdly, we have adopted the Boussinesq approximation to the equations of motion, which formally limits us to a region of small spatial extent near the transition region.

In this framework, we first considered a single `secondary mode' which has the smallest frequency distance from the `primary mode' and formulated a condition for nonlinear effects to be important based on when the amplitude of the near-resonant secondary is comparable to that of the primary (see eq. (\ref{mp15})). Our analytical calculations were supported by a numerical study in \S~\ref{numerical}, where we analysed the generation of secondary waves as a linear problem with a source term determined by the presence of the primary wave, which confirmed our analytical results. 

We discussed how our model problem can be related to the normal modes of a spherical star with a radiative interior and convective envelope in \S~\ref{relation}. We assumed the primary gravity waves were excited in a non-rotating star by a point-like perturber (i.e.~planet or close binary star) on a circular orbit to illustrate this effect. We formulated the criterion (\ref{mp22}), which is based on (\ref{mp15}), and specifies $C_{crit,c}$, which is the smallest value of $q/(1+q)$, where $q$ is the mass ratio (secondary perturber mass/primary star mass), needed for the secondary mode with the closest frequency to (double) the tidal forcing frequency, to exceed the amplitude of the primary tidal wave, which occurs when $q/(1+q)>C_{crit,c}$. We point out later that several modes with frequencies sufficiently close to the tidal frequency could be effectively excited under certain assumptions, and provide the corresponding condition (\ref{mp26}), which can be used if both the secondary modes are non-dissipative and when the condition of ``moderately large dissipation" can be applied to the secondary modes themselves (approximately equivalent to the statement that the mode damping time is shorter than the group travel time for a gravity wave packet across the radiation zone). This condition defines the quantity $C_{crit}^{dense}$, which is analogous to $C_{crit, c}$, but can be applied when there is a sufficiently dense spectrum of secondary modes, and typically gives a slightly more optimistic estimate of nonlinearity. In addition, we formulate a similar criterion for wave breaking near the centre of a star with a fully radiative interior (see eq. (\ref{e12}) and the same condition, but for a star with a convective core near the core itself (see eq. (\ref{core2}) in the formalism of \cite{IPCh}. Both (\ref{e12}) and (\ref{core2}) are analogous to similar criteria reported by \citep[e.g.][]{GD98,OL2007,BO2010,S18,Barker2020}. These conditions are formulated in terms of $C_{crit, centre}$ and $C_{crit, core}$, respectively, and for these nonlinear effects to be important we require $q/(1+q)>C_{crit,centre}$ or $q/(1+q)>C_{crit,core}$, respectively.

We applied our results to stellar models to estimate the importance of the nonlinear excitation of super-harmonic secondary waves in \S~\ref{application}. We speculate that the conditions for transition to non-linear behaviour near radiative-convective interfaces could result in an efficient damping of primary tidal waves, thus justifying the assumption of ``moderately large dissipation". We calculated $C_{crit, centre}$, $C_{crit, core}$, $C_{crit, c}$ and $C_{crit}^{dense}$ for a number of models of main-sequence stars with masses $M=1$, $1.4$ and $2M_{\odot}$ and different ages. We found the condition for wave breaking near the centre is always more important in our solar-mass model (and others with radiative cores) than the condition for nonlinear behaviour near $r_c$, consistent with prior work. The opposite situation occurs however in more massive stars. Apart from two models with extended convective envelopes, and, for one with a radiative centre, for all other models, ages, and orbital periods, $C_{crit,c}$ and the average values of $C_{crit,c}^{dense}$ are smaller, and, in many cases, much smaller than $C_{crit, core}$. In Fig.~\ref{Ct} we show, for example, the behaviour of these quantities for $M=2M_{\odot}$ and a fixed orbital period $P_{orb}=1\mathrm{d}$, with stellar age. This figure shows that for ages of order $10^{9}\mathrm{yr}$, $C_{crit,  c}$ can be as small as $10^{-3}$ (i.e. relevant for planetary-mass companions). If our assumption that the transition to nonlinear behaviour results in efficient dissipation is valid, this implies that the nonlinear generation of super-harmonics by tidally-excited gravity waves could be important for hot Jupiters and other massive companions orbiting these stars.

{In this paper we have ignored stellar rotation (with angular velocity $\Omega_r$) except in Appendix~\ref{app}, under the assumption that the star is rotating slowly relative to the tidal frequency $\omega\approx 2\Omega_{orb}$, i.e. that $\Omega_r^2\ll \omega^2$, and certainly that $\Omega_r^2\ll \Omega_*^2$. The neglect of rotation is likely to be valid for predicting the current and future evolution of most, but not all, of the shortest-period hot Jupiter systems observed. This is because their host stars typically rotate with periods much longer than their planetary orbits \citep[e.g. see the table in Appendix C of][]{Barker2020}, so we can probably neglect the corresponding frequency shifts due to rotation (as long as their radiation zones rotate similarly to their surfaces). For example, WASP-12 is inferred to have a rotation period longer than 23 days and perhaps as long as 38 days \citep[based on the observed $V\sin i$, see e.g.][and references therein]{Patra2020}, whereas the planet WASP-12 b orbits in only 1.09 days. A brief calculation of the possible rotational correction to our results in Appendix~\ref{app} confirms that it doesn't appear to be significant for such orbital and rotational periods. In addition, inertial waves (restored by Coriolis forces) will not be excited by planetary tidal forcing in these stars (which would require $\omega^2\leq 4\Omega_r^2$, in which case we could no longer describe the modes as a single spherical harmonic in the form of equation~\ref{l4}). However, the effects of rotation on the mechanism we have analysed should be explored in more detail in future work to enable us to study more rapidly rotating planetary hosts, such as young stars or those that are nearly rotating synchronously with their planetary orbits e.g. $\tau$-Boo or WASP-128.}

Our initial study here should, however, be considered as a preliminary one. Firstly, we should confirm our speculation that efficient damping of the primary tidal wave occurs by exciting super-harmonic secondary waves using direct nonlinear numerical calculations. Secondly, we should extend our results to spherical geometry, incorporate stellar rotation, as well as more realistic Brunt-V\"{a}is\"{a}l\"{a} frequency profiles near $r_c$, and the possibility of overshooting in the transition region, etc. These issues are left for future work. Finally, 
a different mechanism to efficiently damp gravity waves in stars with convective cores could involve the excitation of a primary gravity mode to large amplitudes via resonance locking \citep[e.g.][]{WS99,WS02,Zanazzi21,Ma21}. This mechanism is unlikely to operate effectively in stars with radiative cores due to the likelihood of wave breaking, as predicted by (\ref{e12}) in this paper, for example, but it could potentially be important in stars with convective cores and should be explored further in those stars. The interaction with nonlinearly generated super-harmonics should also be considered in that problem.

\section*{Acknowledgements}
We would like to thank the referee Michael Efroimsky for a very prompt and useful report. We are also grateful to Y. Lazovik for important remarks. PBI and SVCh were supported in part by the grant 075-15-2020-780 (N13.1902.21.0039) 'Theoretical and experimental studies of the formation and evolution of extrasolar planetary systems and characteristics of exoplanets' of the Ministry of Science and Higher Education of the Russian Federation. AJB was supported by STFC grants ST/R00059X/1 and ST/S000275/1.
 
\section*{Data availability}
There are no new data associated with this article.

\appendix 

\section{The correction to $\nu$ produced by slow rotation}
\label{app}
Rotation also causes a difference between the secondary wave frequency, $\omega_s$, and twice the primary wave frequency, $2\omega_p$. We estimate it below assuming that the rotation axis 
of the star is perpendicular to the orbital plane\footnote{While hot Jupiters may have significant inclinations of their rotational axes with respect to orbital plane, this is less common for the shortest-period planets \citep[e.g.][]{Albrecht2012}.} and the stellar rotational frequency, $\Omega_r$ (assumed constant), is much smaller than the frequencies of interest. We neglect effects that are quadratic in $\Omega_r$, and consider $\omega_s$ and $\omega_p$ as defined in the inertial frame, which is different from our definitions in the main text, where it is implied that these frequencies are defined in the
rotating frame. As discussed in Section \ref{dense}, when the frequencies are considered in the inertial frame the primary frequency is simply double the orbital frequency, $\omega_p=2\Omega_{orb}$. For a high-order gravity mode with a frequency $\omega$ 
the leading order rotational frequency correction is determined by
\citep[see e.g.][Equations 8.45 and 8.46]{CD1998}
\begin{equation}
\omega\approx\omega_0\left(1+\left(1-{1\over L^2}\right){m\Omega_r\over \omega_0}\right),
\label{a1}
\end{equation}
where $\omega_0$ is the mode eigenfrequency for a non-rotating star, and $m$ is the azimuthal wavenumber. We
remind the reader that $L=\sqrt{l(l+1)}$, and we use $l=m=2$ for the primary mode, and $l=m=4$ for the secondary mode. This
choice originates from a comparison of our planar problem with the full spherical problem. 
Similar to what is done in Section \ref{local_spherical}, we calculate the difference $2\omega_p-\omega_s$ using $\omega_0=\omega_{p}$ and $\omega_0=2\omega_p$ for the primary and secondary mode, respectively, to obtain
\begin{equation}
2\omega_p-\omega_s\approx-{7\over 15}\Omega_r.
\label{a2}
\end{equation} 
Comparing (\ref{a2}) with (\ref{mpn3}) we see that slow stellar rotation can be accounted for in the expression for $\nu$ by redefining $f(n)$ entering (\ref{mpn5}) according to 
\begin{equation}
f(n) \rightarrow f(n)-{7\over 60}{\Omega_r\over \Omega_{orb}}.
\label{a3}
\end{equation} 
It is seen from (\ref{a3}) that the additional term is rather small, approximately $3.8\times10^{-3}$, for orbital periods of approximately $1$d and rotational periods of order $30$d, which are appropriate values for many short period hot Jupiters like WASP-12. Comparing this value with typical values of $f(n)$ shown in Fig.~\ref{fn}, we see that this term doesn't appear to be significant 
for such systems. However, it may be important for faster rotators. Finally, note that the correction is negative for prograde rotation, and therefore, the presence of the additional term in $\nu$ could, in principal, make it negative. In this case its absolute value should be used in our criteria for predicting non-linear behaviour.

\label{lastpage}

\end{document}